\documentclass[final,3p]{./elsarticle}
\usepackage{subfigure}
\usepackage{overpic}
\usepackage{graphicx}
\usepackage{gensymb} 
\usepackage{amsmath,amsfonts,amssymb}
\usepackage{overpic}
\usepackage{contour}\contourlength{0.7pt}
\usepackage{algorithmicx}
\usepackage{algorithm}

\begin{document}

\begin{frontmatter}

\title{Smooth quasi-developable surfaces bounded by smooth curves}

\author[a]{Pengbo Bo \corref{pengbo}}
\cortext[pengbo]{Corresponding author}
\ead{pbbo@hit.edu.cn}
\author[a]{Yujian Zheng \corref{yujian}}
\ead{paul.yj.zheng@gmail.com}
\author[b,c]{Caiming Zhang \corref{caiming}}
\ead{czhang@sdu.edu.cn}
\address[a]{School of Computer Science and Technology, Harbin Institute of Technology, Weihai, Weihai 264209, China}
\address[b]{Shandong Co-Innovation Center of Future Intelligent Computing, Yantai 264025, China}
\address[c]{Shandong Province Key Lab of Digital Media Tachnology, Shandong University of Finance and Economics, Jinan 250061, China }

\begin{abstract}
Computing a quasi-developable strip surface bounded by design curves finds wide industrial applications. Existing methods compute discrete surfaces composed of developable lines connecting sampling points on input curves which are not adequate for generating smooth quasi-developable surfaces. We propose the first method which is capable of exploring the full solution space of continuous input curves to compute a smooth quasi-developable ruled surface with as large developability as possible. The resulting surface is exactly bounded by the input smooth curves and is guaranteed to have no self-intersections.  The main contribution is a variational approach to compute a continuous mapping of parameters of input curves by minimizing a function evaluating surface developability.  Moreover, we also present an algorithm to represent a resulting surface as a B-spline surface when input curves are B-spline curves.
\end{abstract}


\end{frontmatter}



\section{Introduction}

Developable surfaces  have a promising  geometric property of accurate isometric mapping into  a plane. Therefore, the developable surface is a frequently used  mathematical model for representing inextensible materials  such as metal sheets and wood panels.
Devalopable surfaces find  wide applications in industrial applications such as architectural free-form surface design and ship-hull design where designing a quasi-developable strip taking two design curves as its boundary curves is a useful tool~\cite{Pottmann2008dstrip}~\cite{Perez2007}. For manufacturing purposes, a developable surface represented as a ruled surface is usually preferred  because  rulings serve as important guidance in a manufacturing process using a folding machine.  Due to its importance, finding a quasi-developable surface in the form of a ruled surface interpolating two specified curves is widely investigated and a variety of approaches have been proposed.

In the discrete setting, where  input curves are polylines, existing methods  search for a sequence of \emph{bridge lines}  of  large developability connecting vertices of the given polylines. However, when the input curves are smooth curves, e.g. B-spline curves, computing a finite set of  lines to form  a discrete surface does not explore the whole continuous solution space. A method making a full use of smooth input curves is required to meet the following requirements.  
\begin{enumerate}
\item A method should be able to search continuously on input curves for developable ruling lines instead of searching in a  finite set of bridge lines.
\item A  quasi-developable surface is required to  be a smooth surface taking input curves as its exact boundaries. 
\end{enumerate}

Using existing methods designed for discrete inputs, the first task can be partially  solved  by taking a large number of sample points on input curves, which however may lead to a considerable increase of computation time ~\cite{Charlie2005Strip}~\cite{wang2005optimal}. For the second task, existing methods employ a standard lofting method to compute a surface interpolating  sample points on input curves ~\cite{Perez2007}, or create a  surface by smoothly stitching together B$\acute{e}$zier surface patches defined between bridge lines ~\cite{Chen2013}.  To the best of our knowledge, no existing methods can compute a smooth ruled surface with as large developability as possible taking two specified smooth curves as its exact boundaries.

We propose a method which directly computes a smooth surface as developable as possible bounded by two input smooth curves. The surface we obtain is a ruled surface which is guaranteed to have no self-intersections.  The main contributions of our work are listed as follows.

\begin{enumerate}
\item We propose the first method  which is capable of exploring the full solution space of  continuous input curves to find a ruled surface with as large developability as possible.
\item The resulting surface takes the input curves as its exact boundaries and is guaranteed  to be free of self-intersections.
\item We also propose an algorithm to write a resulting surface obtained with our method as  a B-spline surface.
\end{enumerate}

\section{Related works}

\emph{Discrete developable surfaces.} Developable surface modeling has been widely studied  in various disciplines.  In developable object simulation, the triangular meshes are the most frequently used representations. English et al. proposed a physically-based method for animating developable surfaces with nonconformal faces ~\cite{English2008Nonconforming}. Solomon et al. encoded a developable surface as a set of ruling lines and found positions of ruling lines by relaxing mean curvature bending  energy ~\cite{Solomon2012Flexible}.
Oded et al. proposed a local operator to modify a triangular mesh into piecewise developales  ~\cite{Stein2018Developbility}. A planar quadrilateral mesh is another type of discrete developable surface which is based on a solid theoretical foundation. Liu  et al. proposed a method for modeling a developable surface  with  a quadrilateral mesh by optimizing face planarization and performing mesh subdivision in an alternative manner ~\cite{Liu2006Conical}. Rabinovich et al. make use of quadrilateral meshes and define a discrete orthogonal geodesic  net to model developable surfaces ~\cite{Rabinovich2018discreteGeodesicNet}.

Triangular meshes are also often employed in industrial  design of developable shapes. Liu et al. propose to approximate a parametric surface with a minimum set of triangle strips with $C^{0}$ continuity ~\cite{LiuYongJin2007DevelopableStrip} ~\cite{LiuYongJin2006Stripification}.  Rose et al. compute piecewise developable triangular meshes from arbitrary design curves ~\cite{Rose2007DevelopableFromSketch}. Liu et al. invented a system for interactive design of   developable triangular  meshes from sketched curves using a touching panel device ~\cite{Liu2011IndustrialDesign}.  These methods are not capable of creating smooth developable surfaces. 

\emph{Smooth developable surfaces.} Pottmann et al. use a composition of B-spline developable strips to approximate freeform architectural models  ~\cite{Pottmann2008dstrip}, which essentially solve a constrained B-spline surface fitting problem~\cite{Bo2012sdm}. Tang et al. proposed an interactive developable surface design approach to create composite B-spline developable surfaces ~\cite{tang2016interactive}, in which a shape is designed through an incremental procedure of  user modification and surface optimization. 
Bo et al. make use of the rectifying developable and propose an algorithm for modeling smooth developable surfaces from geodesic curves on surfaces ~\cite{Bo2007Geodesic}. Complex developable surfaces are modeled by composing cone patches with smooth transitions ~\cite{Hwang2015WrapCone}.  These design methods cannot be applied directly to compute a developable surface bounded by specified curves.

\emph{Developable surfaces bounded by specified curves.}  A particular problem in industrial design is modeling developable strips bounded by design curves. This problem is frequently encountered in fabrication with inextensible materials ~\cite{Jung2005SketchFold}.  One example is the ship-hull design with a few dominating feature curves as input and developable surface patches are computed to interpolate the feature curves ~\cite{Perez2007}. Other industrial applications include shoe design and garment design. When the input curves are polylines, a developable mesh surface is created.  Tang and Wang ~\cite{tang2005modeling} simulated the unfolding process of a bendable sheet to generate a bridge triangulation of vertices of input polylines to form a developable surface. In order to find a global optimum in the solution space of all reasonable  triangulations, a developable triangulation  problem is formulated as a graph problem and the Dijkstra's algorithm is utilized to solve it ~\cite{wang2005optimal}. 
Chen et al. proposed a local-global method which allows a perturbation of mesh vertices to optimize surface developability~\cite{chenming2011}.

To create  smooth developable strips, some existing methods compute a developable surface to satisfy particular geometric constraints such as  interpolating specific curves or rulings. Both the B$\acute{e}$zier surface and the B-spline surface are considered as representations for developable surfaces. Accurate developable surfaces are obtained by solving  nonlinear equations of various developability constraints ~\cite{Pottmann1995Rational}~\cite{Chenal1999approximation}~\cite{Aumann2003Simple}~\cite{Aumann2004DegreeElevation}.  The dual representation of a developable surface is also considered which however is not intuitive for shape control ~\cite{Pottmann1999Approximation}.  These methods compute accurate developables and do not work for  modeling developable surfaces bounded by two arbitrary  curves. In ~\cite{Perez2007}, quasi-developable B-spline strips are created for ship-hull design using ruling searching and surface lofting. The multi-conic method is used to modify the shape of feature curves to improve surface developability. Subag and Elber approximate a freeform NURBS with piecewise developable surfaces with a global error bound ~\cite{Subag2006Piecewise}. Chen and Tang improved the method in ~\cite{wang2005optimal} by stitching smooth surface patches to obtain  a smooth surface. Both  triangular  B$\acute{e}$zier patches with $G^{1}$ continuity~\cite{chen2013quasi}  and quadrilateral  B$\acute{e}$zier patches with $G^{2}$ continuity ~\cite{Chen2013} are studied.   With these methods, a smooth surface is computed based on the discrete ruling lines and  the generated quasi-developable surface interpolates some data points of the input curves.  They are however not capable of generating a smooth quasi-developable  ruled surface taking  the input smooth curves as its exact boundary curves.

We are particularly interested in modeling smooth quasi-developable strips exactly bounded by specified  smooth design curves.  Due to the fact that NURBS surfaces are industrial standard representations for surfaces, it is desired that a resulting surface can be represented as a B-spline surface. We propose algorithms to fulfill these tasks which cannot be solved with existing methods.

\section{Developable surfaces from parametric curves}

We aim to  find a smooth developable strip bounded by two parametric curves, i.e. $C_1(t)$, $t\in[0,1]$  and $C_2(T)$, $T\in[0,1]$. In our work, we use degree $p$ B-spline curves for the input curves.  It is clear that there are in general no accurate developable  surfaces interpolating two arbitrary curves unless the curves lie exactly on a developable surface. Therefore, in industrial applications,  computing a surface strip  which is with as large developability  as possible is desired. 
We represent a developable surface as a ruled surface composed of a family of infinite lines, i.e., \emph{ruling lines},  connecting points on input curves. Because a ruling line is uniquely expressed by a pair of parameters $(t,T)$ of  input curves,  a continuous  surface  is uniquely defined by  a continuous function of parameter mapping $T=\sigma(t)$   to be an infinite set of ruling lines   $\overline{C_1(t)C_2(\sigma(t))}$.   Such a function $\sigma(t)$ is called a parametrization mapping function in ~\cite{wang2005optimal}.  Note that the end ruling lines, i.e. $\overline{C_1(0)C_2(0)}$ and $\overline{C_1(1)C_2(1)}$, are in general not developable and  curve extension as a preprocessing is needed in some cases as explained in Section ~\ref{sec:initialization}.

Existing methods take some sample points on input curves and compute a developable triangulation of the points, or search for some bridge lines of large developability connecting  the sample points. The bridge lines, in fact, define samplings of some continuous parametrization mapping function.  This kind of strategy is  not capable of searching on a continuous curve  to explore an infinite number of samplings. Therefore, a large  number of sampling points are needed to improve the possibility of finding a high degree of developability, which however may lead to a large increase in computation time. Our experiments show that a tiny difference in parametrization mapping may result in a considerable difference in developability of ruling lines, as is demonstrated by  an experiment in Figure ~\ref{fig:degree2}.  Therefore,  a capability of continuous searching on a continuous curve is extremely important  for  achieving as large developability as possible.

\subsection{A variational framework for computing a continuous parametrization mapping}
\label{sec:variarional_framework}
To provide with a capability of searching in a continuous parameter domain, we take a non-decreasing sequence of fixed  parameters $t_i$,$i=0,...,K$ on one  curve $C_1(t)$ and search for  corresponding parameters $T_i$, $i=0,...,K$ of the other curve $C_2(T)$ in the continuous domain $[0,1]$, such that the ruling lines $\overline{C_1(t_i)C_2(T_i)}$ achieve large developability. Let $E(t_i,T_i) = 0 $ be a constraint of surface developability employed on a ruling line $\overline{C_1(t_i)C_2(T_i)}$, the solution to the  minimization problem  in Eqn.(\ref{eqn:minproblemT}) corresponds to a sequence of lines $\overline{C_1(t_i)C_2(T_i)}$, $i=0,...,K$ with large developability.  

\begin{equation}
\label{eqn:minproblemT}
\min{\sum_{i=0}^{K}E(t_i, T_i)},
\end{equation}
where $E(t_i, T_i)$ is a function evaluating surface developability at a ruling line $\overline{C_1(t_i)C_2(T_i)}$.

In order to avoid intersections of rulings lines,  a sequence  of scalar values $T_i$, $i=0,...,K$  is required to be monotonically increasing. This requirement can be met by defining $T_i= \sum_{j=0...i}{(\alpha_j)^2}$, $i=0,...,K$ to ensure that the relationships $T_0 \geq 0$, $T_{i+1} \geq T_{i}$, $i=0,...,K-1$ are always satisfied. The parameters $\alpha_j$, $j=0,...,K$  serve as unknown variables in the objective function in Eqn.(\ref{eqn:minproblemT}).

Solving the minimization problem in Eqn.(\ref{eqn:minproblemT}) enables us  to perform a continuous search on $C_2(T)$ for rulings with large developability. As a result, we obtain a sequence of ruling lines and  a discrete surface as a quadrilateral mesh is directly obtained. This method is called a \emph{discrete mapping method} because it does not produces a continuous mapping function.  When a continuous surface is  demanded,  a continuous function $c(t)$ is computed to  meet $c(t_i)=T_i$, $i=0,...,K$, which defines a continuous surface $S$ bounded by $C_1(t)$ and $C_2(T)$. However, besides the ruling lines from the solution to the minimization problem in Eqn.(\ref{eqn:minproblemT}), surface regions on $S$ are defined by the rulings from function interpolation whose developability is not guaranteed. The experiment in Figure ~\ref{fig:Helical} illustrates this fact. Increasing the number of sample parameters can partially solve this problem, with the cost of  an increase in the number of unknown variables in optimization and computation time.

In order to provide a large number of controls over a mapping function without introducing a large number of variables in optimization, we use a continuous function $\sigma(t)$ with a specific type for  parametrization mapping.  A function $\sigma(t)$ which makes $S$ achieve a large developability is  obtained   by   minimizing a function evaluating surface developability at some sampling ruling lines. That is, we solve the following minimization problem for  shape parameters in   $\sigma(t)$.

\begin{equation}
\label{eqn:minproblem}
\min{\sum_{i=0}^{K}E(t_i,\sigma(t_i))},
\end{equation}
where $E(t_i, \sigma(t_i))$ is a function evaluating surface developability at a ruling line $\overline{C_1(t_i)C_2(\sigma(t_i))}$, where $K$ can be much larger  than the number of unknown variables in optimization.   In order to obtain a surface without self-intersections, $\sigma(t)$ is required to be monotonically increasing for which we use a B-spline function as discussed in the following section.

\subsection{B-spline parametrization mapping function}

An optimal parametrization mapping function $\sigma^{*}(t)$ which achieves the largest developability is, in general, a highly nonlinear function and  hence only an  approximation to it can be computed. Solving the minimization problem in Eqn.(\ref{eqn:minproblem})  drives the function values $\sigma(t_i)$ to meet the values of the unknown function  $\sigma^*(t_i)$, i.e., to minimize $\Vert \sigma^*(t_i) -\sigma(t_i) \Vert$.  Therefore, the performance of our approach depends on the approximation ability of the employed function.   We use a degree $d$ B-spline function defined in Eqn.(\ref{eqn:bsplinefunction}), which is essentially a piecewise polynomial function, for its proven ability  of approximating an arbitrary function. 

\begin{equation}
\label{eqn:bsplinefunction}
\sigma(t) =\sum_{i=0}^{n}{\beta_i B_{i,d}(t)},
\end{equation}
where  $\beta_i$ act as shape parameters; $B_{i,d}(t)$ are degree $d$ B-spline basis functions. A clamped knot vector $[t_0=t_1=...=t_{d+1}=0,t_{d+2}, ..., t_{n},t_{n+1}=...=t_{n+d+1}=1]$ is used.  To make  $\sigma(t)$  a monotonically increasing function, we observe the first derivative of $\sigma(t)$ in Eqn.(\ref{eqn:firstder}) and conclude that  $\beta_i \geq \beta_{i-1}$, $i=1,...,n$ are sufficient conditions to make $\sigma(t)$ in Eqn.(\ref{eqn:bsplinefunction})  a monotonically increasing function. 
\begin{equation}
\label{eqn:firstder}
\sigma'(t)= d\sum_{i=0}^{n}{\frac{\beta_i-\beta_{i-1}}{t_{i+d}-t_{i}}B_{i,d-1}(t)}.
\end{equation}

Therefore, we define a function of a specific form by enforcing   $\beta_{i+1}=\beta_i+\varepsilon_{i}^{2}$, $i=0,...,n-1$. We also enforce the first coefficient  to be a non-negative value by defining $\beta_0=\varepsilon_0^{2}$. In summary, a B-spline function defined in  Eqn.(\ref{eqn:mappingfunction}) is employed for parametrization mapping, which is quadratic in the shape parameters $\varepsilon_i$, $i=0,...,n$.

\begin{equation}
\label{eqn:mappingfunction}
\sigma(t)=\sum_{i=0}^{n}{ \left( \sum_{j=0}^{i}\varepsilon_{j}^{2} \right) B_{i,d}(t) }
\end{equation}

Using a B-spline mapping function has another virtue of representing the final quasi-developable surface as a B-spline surface using an algorithm presented in Section~\ref{sec:bspline_representation}.

\subsection{Computing the parametrization  mapping function}

We  aim to find a function $\sigma(t)$ to approximate the unknown function $\sigma^*(t)$ of optimal parametrization mapping, by solving the minimization problem in Eqn.(\ref{eqn:minproblem}), where $\sigma(t)$ is a B-spline function defined in Eqn.(\ref{eqn:mappingfunction}). The developability constraint is expressed by  $E(t_i,\sigma) = 0$  in which $E(t_i, \sigma)$ is a function evaluating surface developability at a ruling line $\overline{C_1(t_i)C_2(\sigma(t_i))}$.

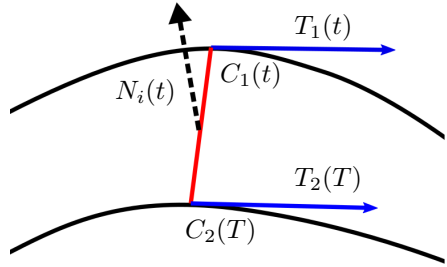
\begin{figure}[!ht]
\centering
\begin{overpic}[width=0.35\columnwidth]{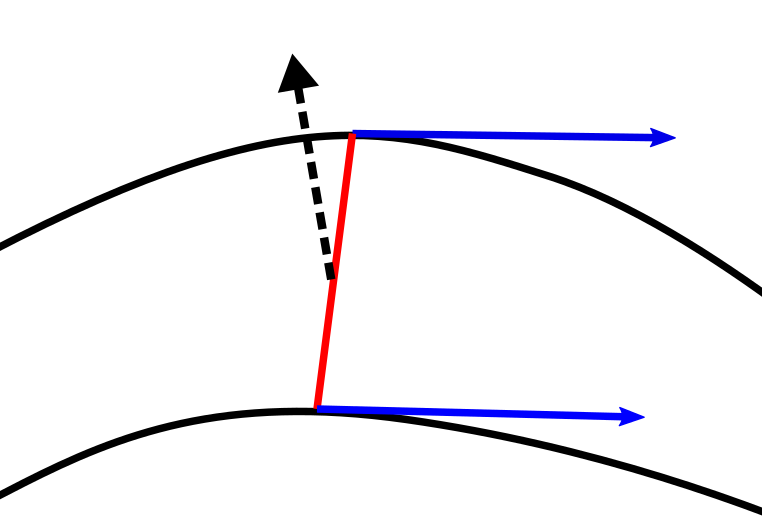}
  \put(65,55){$T_1(t)$}
  \put(65,20){$T_2(T)$}
   \put(24,40){$N_i(t)$}
     \put(48,44){$C_1(t)$}
  \put(40,8){$C_2(T)$}
\end{overpic}
\caption{Developability constraint.}
\label{fig:DevelopableConstraint}
\end{figure}

It is well-known that  a ruling line is developable if it has an identical normal vector along it, which is  equivalent to the coplanarity of three vectors expressed by
 \begin{equation}
 \label{eqn:developable_condition}
  E_1(t_i, \sigma):=det(C_1(t_i)- C_2(\sigma(t_i)),   C_1'(t_i), C_2'(\sigma(t_i)))=0, 
 \end{equation}
where the comma symbol denotes the first derivative to $t$. See Figure ~\ref{fig:DevelopableConstraint}.

The developability condition of Eqn.(\ref{eqn:developable_condition}) involves in a  degree $4p$ polynomial in terms of the shape parameters $\varepsilon_i$ of the mapping function  $\sigma(t)$. Recently, Tang et al. proposed an equivalent condition to Eqn.(\ref{eqn:developable_condition}) by introducing a vector function of  normal spline $N(t)$ represented by a B-spline curve ~\cite{tang2016interactive}, which can be  expressed as 
\begin{equation}
N(t) \cdot C_1'(t)= N(t) \cdot C_2'(\sigma(t))= N(t) \cdot L(t) =0,
\end{equation}

where  $L(t):=C_1(t)- C_2(\sigma(t))$ is a ruling vector which involves in the shape parameters of $\sigma(t)$.  See Figure ~\ref{fig:DevelopableConstraint} for an illustration. This new condition consists of some equations of degree $2p+1$ at most and is adopted in our work.  However, a smooth normal spline curve  possesses an inherent fairness with it which may prevent us from obtaining a high degree of developability. Therefore, we use normal variables $N_i$, $i=0,...,K$ instead of $N(t)$ in optimization which is a special piecewise linear B-spline curve. A  constraint  relating to a sampling $\sigma(t_i)$ is hence defined by 

\begin{equation}
E_2(t_i,\sigma) := (N_{i} \cdot C_{1}'(t_i))^{2} + (N_{i} \cdot C_2'(\sigma(t_i)))^{2} + (N_{i} \cdot L(t_i))^{2} =0.
\end{equation}

The computation of $\sigma(t)$ is formulated as an optimization problem of minimizing a function evaluating surface developability  at  some samplings of  $\sigma(t)$.  The objective function is defined by

$$F_{Dvlp} := \sum_{i=0}^{K} \left( (N_{i} \cdot C_{1}'(t_i))^{2} + (N_{i} \cdot C_2'(\sigma(t_i)))^{2} + (N_{i} \cdot L(t_i))^{2}  \right).$$

To avoid degeneration of the normal variables, we use a regularization term  
\begin{displaymath}
F_{Reg} := \sum_{j=0}^{K} (\Vert N_{j} \Vert^{2}-1)^{2}.
\end{displaymath}

We, therefore, solve the following minimization problem for a parametrization mapping function $\sigma(t)$.

\begin{equation}
\label{eqn:objfun}
\min\limits_{P,N}{\lambda_1 F_{Dvlp} +   \lambda_2 F_{Reg}}, 
\end{equation}
where $P$ and $N$ are variables in optimization with $P$ denoting a set of shape parameters  of  $\sigma(t)$: $P=\{ \varepsilon_i , i=0,...,n\}$; $N=\{ N_i, i=0,...,K\}$ being  a set of normal variables.

 The minimization problem is solved with the L-BFGS algorithm  for which only the first derivative to unknown variables are needed which are computed explicitly ~\cite{Zheng2012LBFGS}.  All models are scaled into  a unit box to make the adjustment of the coefficients convenient. The weights to the  terms in the objective function in Eqn.(\ref{eqn:objfun}) are set as empirical values. For all experimental models in this paper, we use $\lambda_1= 100$,  $\lambda_2=1$.  Figure ~\ref{fig:degree2} shows an example in which the input  curves defines an exact developable surface with a known parametrization mapping function. It demonstrates that our method is able to compute a parametrization mapping function which approximates the accurate mapping function in high precision.

\subsection{Initialization of parametrization mapping function} 
\label{sec:initialization}

Our algorithm  needs an initial parametrization mapping function as a start of optimization. A good initialization  can  often speed up the convergence of a local optimization method. An initial mapping function can be computed using a sequence of bridge lines $L=\{L_i,i=0,...,m\}$ with large developability.  Each bridge line $L_i$ corresponds to a function value $T_i$  associated with a parameter $t_i$. A  B-spline function interpolating or approximating these values  is computed  as  an initial mapping function. In order to get a monotonic function, a monotonic preservation curve interpolation or approximation algorithm should be employed ~\cite{Wolberg1999monotonic}~\cite{andersson1991monotonic}. A much simpler initialization method is employing  the function $T=t$. 

\begin{figure}[t]
\centering
\begin{minipage}[b]{0.9\linewidth}
\centering
\centerline{
\subfigure[]{
\centering     
\includegraphics[width=0.3\textwidth]{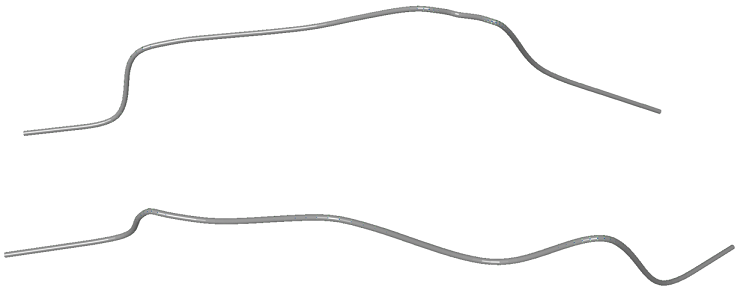}  
}
\hfill
\subfigure[]{
\centering     
\includegraphics[width=0.3\textwidth]{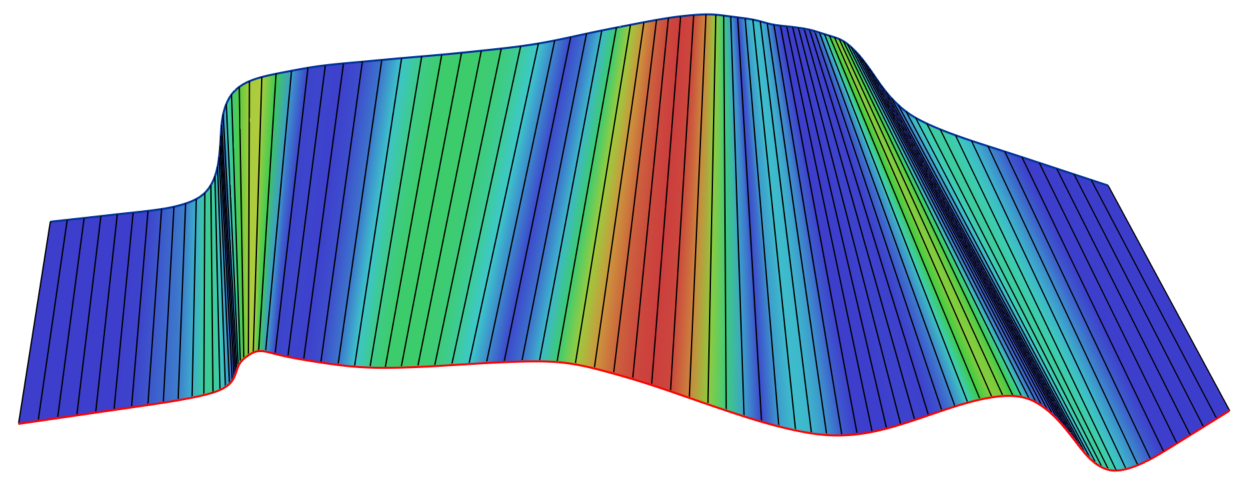}%
}
\hfill
\subfigure[]{
\centering     
\includegraphics[width=0.3\textwidth]{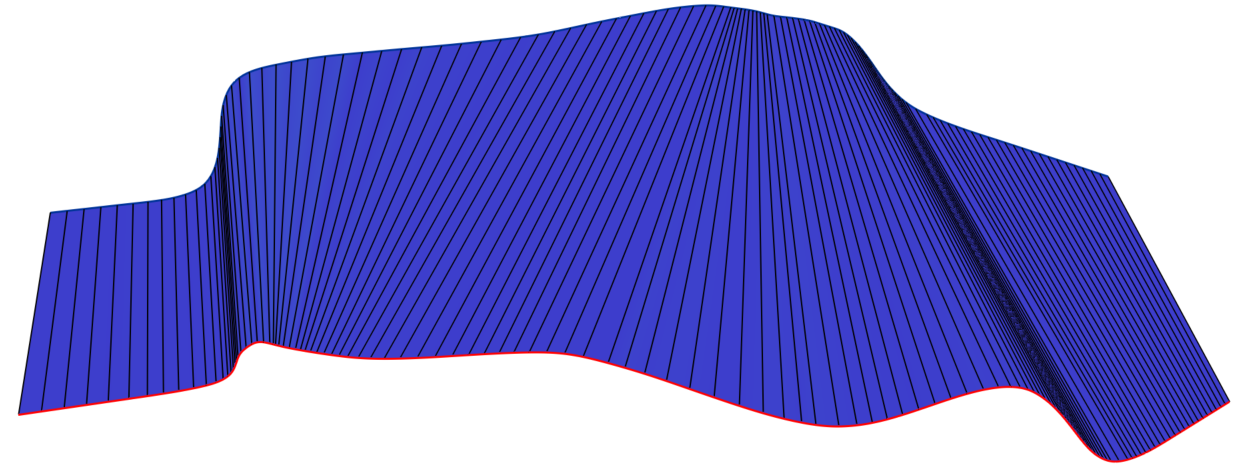}
}}
\end{minipage}
\vfill
\begin{minipage}[b]{0.9\linewidth}
\centering
\centerline{
\hfill
\subfigure[]{
\centering     
\includegraphics[width=0.3\textwidth]{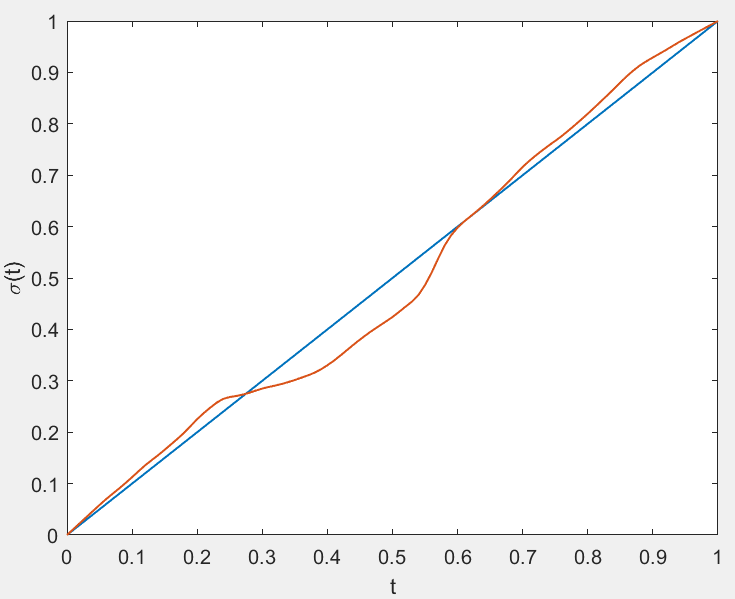}
}\hfill
\subfigure[]{
\centering     
\includegraphics[width=0.3\textwidth]{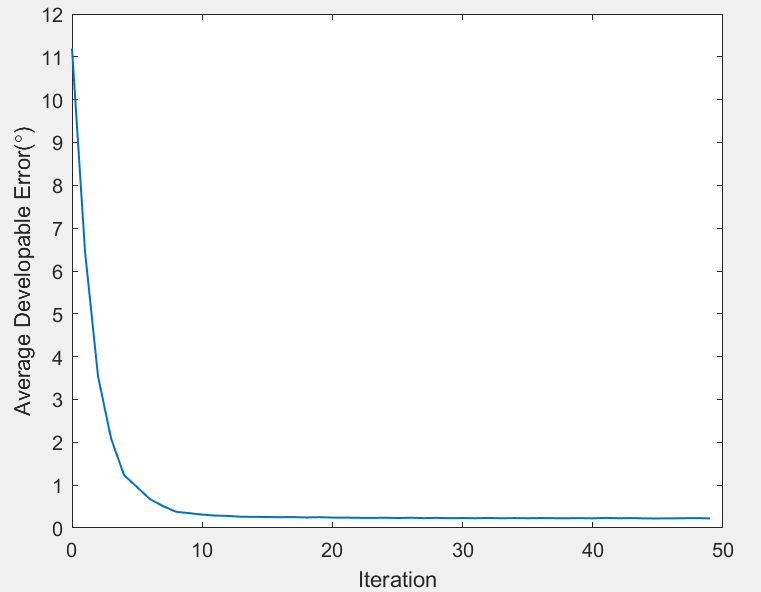}
}\hfill
\subfigure[]{
\centering     
\includegraphics[width=0.3\textwidth]{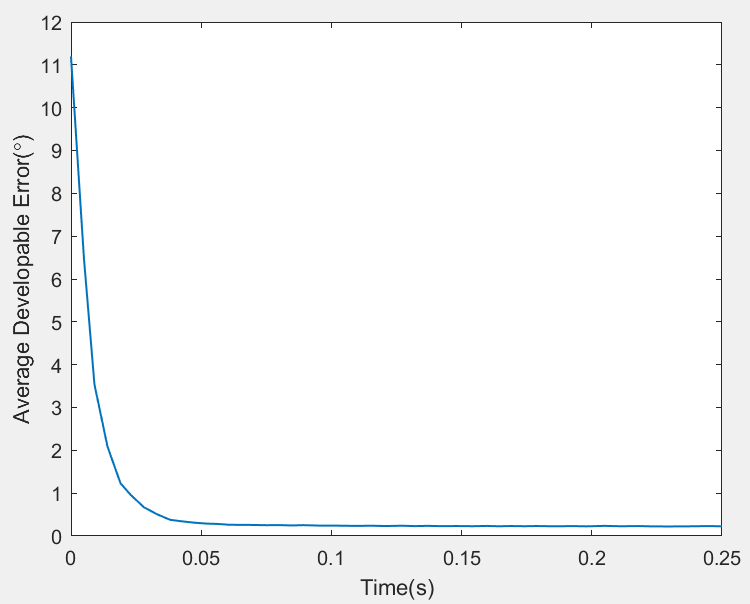}
}\hfill
}
\end{minipage}
\caption{An experiment on  the convergence behaviour of our method. (a) Input curves. (b) Initial ruled surface by the simple initialization. The (maximum,average) warp angle is $(49.1\degree,11.1\degree)$. (c) Result by our method. The (maximum,average) warp angle is $(1.5\degree,0.18\degree)$. (d) Parametrization mapping functions before and after  optimization, where the blue curve refers to the initial function and the red curve refers to the optimized function. (e) Average warp angle vs. iteration. (f) Average warp angle vs. time. }
\label{fig:Fold}
\end{figure}

We implemented both methods and observe that a function initialized by a sequence of developable lines is closer to the final optimized function than using the simple initialization. However,  the procedure of finding some bridge lines with large developability takes a relatively long time compared to the procedure of our optimization algorithm. Our method  finds a solution efficiently even  with the simple function $T=t$ as an initialization. See Figure ~\ref{fig:Fold}  for a demonstration, where the convergence rate of optimization is super-linear and it converges after about 15 iterations which take 0.1 seconds. For all examples shown in this paper, we use the simple initialization method by taking a number of values on $T=t$ as control coefficients.

\begin{figure}[!ht]
\centering
\vfill
\begin{overpic}[width=0.38\columnwidth]{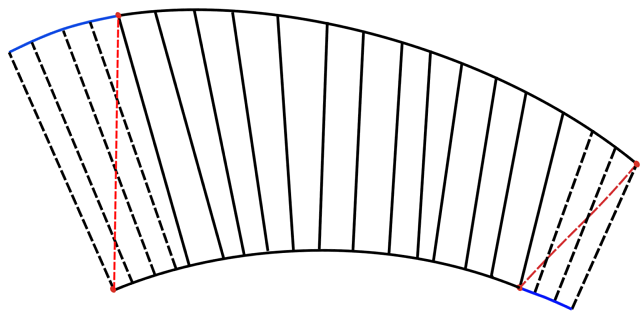}
  \put(96,30){$C_2(1)$}
   \put(72,0){$C_1(1)$}
     \put(15,50){$C_2(0)$}
   \put(3,0){$C_1(0)$}
\end{overpic}
\caption{Curve extension. The curves in black are the original curves; the curve parts in blue are extension parts. The surface regions bounded by the extended curve parts are trimmed by the red lines to get a final result. }
\label{fig:curveExtension}
\end{figure}
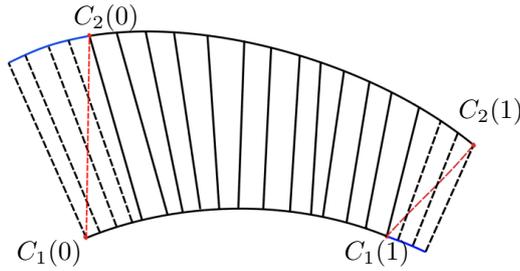

\emph{Curve extension.} It is often the case that the lines connecting the endpoints of the input curves, i.e., $\overline{C_1(0)C_2(0)}$ and $\overline{C_1(1)C_2(1)}$, are not developable, even if the input curves lie exactly on a developable surface. To solve this problem,  curve extension are needed  to deal with two cases. 
\begin{itemize}
\item[-] Case 1.  The obtained mapping function $\sigma(t)$, $t\in[0,1]$ does not cover the whole part of $C_2(T)$. One example of this case is shown by the right part of the surface in Figure~\ref{fig:curveExtension} where we have $\sigma(1) < 1$ before curve extension. In this case, we need to extend  the right end of $C_1(t)$. 

\item[-] Case 2. $C_2(T)$ is not long enough to obtain a surface with large developability. The left part of a surface in Figure ~\ref{fig:curveExtension} belongs to this case.  In this case,  the left end of $C_2(T)$ needs to be extended.
\end{itemize}

In the discussion of our algorithm,  we assume that $C_1(t)$ and $C_2(T)$ have already been extended a bit such that the optimal parameter mapping $\sigma(t)$, $t\in[0,1]$ covers the part on $C_2(T)$ corresponding to the original curve. Obviously, the shape of an extension curve has an impact on  develoapbility of the resulting surface. For curve extension, we  take an extra  point relying on user experience  and utilize the  algorithm in ~\cite{Hu2002} to extend a curve to the extra point.  
ections is demanded. 

Using a parametrization mapping function $T=\sigma(t)$ as a reparametrization of $C_2(T)$, we have a reparameterized curve $\widetilde{C}_2(t)=C_2(\sigma(t))$.  A continuous ruled surface bounded by input curves is thus defined  by

\begin{equation}
\label{eqn:S1}
S(s,t) = C_1(t)(1-s) + \widetilde{C}_2(t)s, s \in[0,1], t\in[0,1],
\end{equation}
which is fully determined if the mapping function $\sigma(t)$ is decided.  It is highly desired that a surface in Eqn.(\ref{eqn:S1}) defined by a B-spline mapping function can be converted into a B-spline surface, while preserving parameter mappings  of input curves.

\section{B-spline quasi-developable surfaces bounded by B-splines curves}
\label{sec:bspline_representation}

In order to convert a ruled surface $S$ defined in Eqn.(\ref{eqn:S1}) into a B-spline surface, we  show an algorithm  to represent $S$ as  piecewise B$\acute{e}$zier surfaces while preserving parametrization mapping  of input curves.  The main steps of  the algorithm is given in Algorithm~\ref{alg:conversion}.

\begin{algorithm}
\caption{Representation conversion into piecewise B$\acute{e}$zier surfaces}\label{alg:conversion}
\begin{algorithmic}[]
\State \textbf{Input.}  Input curves $C_1(t)$ and $C_2(T)$. They are both defined by degree $p$ B-spline curves with $n+1$ control points, with the same clamped knot vector  denoted by $\Gamma_1$ and $\Gamma_2$,respectively, and a parameter domain $[0,1]$.  A degree $d$ B-spline parametrization mapping function $T=\sigma(t)$, $t \in [0,1]$ with $m+1$ control coefficients  and a clamped knot vector $\Gamma_{\sigma}$.  
Note that $\Gamma_{\sigma}$ is in general not equal to $\Gamma_1$ or $\Gamma_2$. 
\State \textbf{Step 1. Knot refinement.} Refine the knot vectors of $\sigma(t)$, $C_1(t)$ and $C_2(T)$ by employing the standard knot insertion algorithm, such that the knot vectors of $\sigma(t)$, $C_1(t)$  are equally expressed by $[t_0,...,t_M]$ and $[\sigma(t_0),...,\sigma(t_M)]$ is the knot vector of $C_2(T)$.
\State \textbf{Step 2. Conversion into B$\acute{e}$zier forms.}  Write each  part  of $\sigma(t)$, $C_1(t)$ and $C_2(T)$, associating to a knot interval of the curve,  in a standard  B$\acute{e}$zier form.
\State  \textbf{Step 3. Reparametrize $C_2$.}  Reparametrize each  B$\acute{e}$zier piece on $C_2$ into a degree $pd$ B$\acute{e}$zier curve  by employing the parametrization mapping function. 
\State  \textbf{Step 4. Elevate the degree of $C_1$. } Elevate each  B$\acute{e}$zier piece on $C_1$ into a degree $pd$  B$\acute{e}$zier curve. 
\State  \textbf{Step 5. Output.} Write the surface in  Equation (\ref{eqn:S1}) as a composition of  B$\acute{e}$zier surfaces of degree $1 \times pd$. 
\end{algorithmic}
\end{algorithm}

The steps in the algorithm are explained  in details in the following. 

\begin{enumerate}
\item[--] \textbf{Step 1.} Let $\Gamma_2^{-1}= \{\sigma^{-1}(T_i), T_i \in \Gamma_2 \}$. The knot vector $\Gamma_{\sigma}$ is refined by inserting the values  $\varphi \in \Gamma_2^{-1} \cup \Gamma_1$ into $\Gamma_{\sigma}$ if $0 < \varphi <  1$ and  $\varphi$ does not introduce any knot duplication.  The obtained knot vector after refinement is denoted by $\Gamma_{\sigma}^{*}$.  

$\Gamma_1$ is refined by inserting the values $v \in \Gamma_{\sigma}^{*}$ into $\Gamma_1$  if the insertion of $v$ does not introduce any knot duplications. The  knot vector after refinement is denoted by $\Gamma_{1}^*$. 

$\Gamma_2$  is refined by inserting the  values $\sigma(v)$ into $\Gamma_2$ if $v \in \Gamma_{\sigma}^*$ and the insertion of $\sigma(v)$ does not introduce any knot duplications. The  knot vector after refinement is denoted by $\Gamma_2^*$.   

After knot vector refinement,  $C_1(t)$  has the same knot vector  as $\sigma(t)$, i.e., $\Gamma_1^*=\Gamma_{\sigma}^*$. Moreover, the values $v_i \in \Gamma_{\sigma}^*$ have a one to one correspondence with the  values $\sigma(v_i) \in \Gamma_2^*$.   

\item[--] \textbf{Step 2.} By employing the knot insertion algorithm to make all internal knots have a duplication equal to the  curve degree, one can convert a B-spline curve with a clamped knot vector into piecewise B$\acute{e}$zier curves ~\cite{PiegTill96TheNurbsBook}.  The control points of every piece of B$\acute{e}$zier curves  are given by the standard knot insertion algorithm.  Once $\sigma(t)$, $C_1(t)$ and $C_2(T)$ are converted into piecewise B$\acute{e}$zier curves in this way, we can write out the B$\acute{e}$zier form of each polynomial piece.

The $j$th polynomial part of  $\sigma(t)$, $t \in[a,b]$ is written  in the B$\acute{e}$zier form  

\begin{equation}
\label{eqn:bezierpart}
\widehat{\sigma}^{j}(\widehat{t})=\sum_{i=0}^{d}{P_i^{(j)}B_{i,d}(\widehat{t})}, \widehat{t} \in[0,1],
\end{equation}
where $P_{i}^{(j)}$ denote control points  of the B$\acute{e}$zier curve.   This step introduces a curve reparamatrization $\widehat{t}=\frac{t-a}{b-a}$ which can be obtained by observing

\begin{equation}
\label{eqn:bezier_repara}
\widehat{\sigma}^{j}(\widehat{t})=  \sigma(\widehat{t}(b-a)+a)= \sigma(t),\widehat{t} \in [0,1].
\end{equation}

Similarly, the $j$th polynomial  curve $C_1(t)$,$t \in [a,b]$ is also written as a standard  B$\acute{e}$zier form   $\widehat{C}_1^{j}(\widehat{t})$, $\widehat{t} \in [0,1]$, with the same  reparametrization  $\widehat{t}=\frac{t-a}{b-a}$. 
The $j$th polynomial piece of $C_2(T)$, $T \in[\sigma(a), \sigma(b)]$ is written as a standard B$\acute{e}$zier curve $\widehat{C}_2^{j}(\widehat{T})$, $\widehat{T} \in [0,1]$ with a reparametrization $\widehat{T}=\frac{T-\sigma(a)}{\sigma(b)-\sigma(a)}$.

\item[--] \textbf{Step 3.} Since the conversion of each polynomial piece into a  B$\acute{e}$zier form introduces reparametrizations, a new  parametrization mapping function of $\widehat{T}$ and $\widehat{t}$ for each piece should be derived. By substituting  $\widehat{T}=\frac{T-\sigma(a)}{\sigma(b)-\sigma(a)}$,  $\widehat{t}=\frac{t-a}{b-a}$ into  $T=\sigma(t)$ and considering Eqn.(\ref{eqn:bezier_repara}),  we have 
\begin{equation}
\label{eqn:repara_mapping}
 \widehat{T}=\frac{\widehat{\sigma}^{j}(\widehat{t})-\sigma(a)}{\sigma(b)-\sigma(a)}, \widehat{t} \in [0,1].
\end{equation}

Substituting  Equation(\ref{eqn:repara_mapping}) into the corresponding B$\acute{e}$zier curve  $\widehat{C}_2^{j}(\widehat{T})$ gives a degree $pd$ polynomial curve parametrized by $\widehat{t}$,  which can be written as a  B$\acute{e}$zier curve  whose control points are  $P_0^{j,2},...,P_{pd}^{j,2}$. This can be realized by using the formula of conversion from power basis to Bernstein basis given in Eqn.(\ref{eqn:basischange}). 

\begin{equation}
\label{eqn:basischange}
t^{k}=\sum_{i=0}^{D}{b_{i,k}B_{i,D}(t)}, 0 \leq  k \leq  D,
\end{equation}
where $b_{i,k}= \frac{C_i^k}{C_D^k}$,  $B_{i,D}$ is a degree $D=pd$ Bernstein basis. $b_{i,k}$ is defined to be zero if $i < k$.  

%

\item[--] \textbf{Step 4.}  The $j$th B$\acute{e}$zier  curve on $C_1$,  denoted by  $\widehat{C}_1^{j}(\widehat{t})$, $\widehat{t} \in [0,1]$, is   elevated into a degree $pd$ curve as follows. We first rewrite $\widehat{C}_1^{j}(\widehat{t})$ in a form  based on the power basis  
\begin{equation}
\label{eqn:curve_power}
\widehat{C}_1^{j}(\widehat{t}) = \sum_{i=0}^{p}f_i \widehat{t}^i + \sum_{j=p+1,...,pd} f_j 0, 
\end{equation}
where $\widehat{t}^i$  means power($\widehat{t}$,$i$), which  is  written as a standard B$\acute{e}$zier curve of degree $pd$ with its  control points   $P_0^{j,1},...,P_{pd}^{j,1}$, using again the formula given in Eqn.(\ref{eqn:basischange}). 

\item[--] \textbf{Step 5.} A standard  B$\acute{e}$zier surface of degree $1\times pd$ for the $j$th piece  is hence written as 
$$ S^{j}(s,t) = \sum_{k=0}^{1}{\sum_{i=0}^{pd}{P_i^{j,k}B_{i,pd}(t)}B_{k,1}(s)},  s \in[0,1], t \in[0,1], $$ 
where $B_{i,d}(t)$ denotes a degree $d$ Bernstein basis function.
\end{enumerate}

The  surface  we obtain is composed of  a sequence of B$\acute{e}$zier surfaces $S^{j}$, $j=0,...,h$ which can be written as a B-spline surface. The control points of a B-spline boundary curve  of this surface corresponding to $C_{k}$, $k=0,1$ are 
$$P^{k}=[P_i^{k}, i=0,...,H \equiv hpd+pd]=[P_0^{0,k}...P_{pd}^{0,k}P_{1}^{1,k}...P_{pd}^{1,k}...P_{1}^{h,k}...P_{pd}^{h,k}],$$
which consist of the control points of the B$\acute{e}$zier curves obtained in the Step 3 or Step 4. Two consequential  B$\acute{e}$zier curves share a common control point which counts only once in the control point sequence of a  B-spline curve. 

Let $\Omega_{t}$ be a knot vector in which any internal knot in $(0,1)$ has a duplication of $pd$ and the end knots, i.e. 0,1, both have a duplication of $pd+1$. Let $\Omega_{s}=[0,0,1,1]$. Then a B-spline surface defined with the knot vectors $\Omega_{t}$, $\Omega_{s}$, associating to the parameters $t$ and $s$ respectively, is written as 

\begin{equation}
S(s,t)= \sum_{k=0}^{1}\sum_{i=0}^{H}{P_{i}^{k}N_{i,pd}(t)N_{k,1}(s)}, s \in[0,1], t \in[0,1],
\end{equation}
where $N_{i,d}(t)$ denotes a degree $d$ B-spline basis function.  

\begin{figure}[t]
\centering
\begin{minipage}[b]{0.9\linewidth}
\centerline{
\subfigure[]{
\centering     
\includegraphics[width=0.3\textwidth]{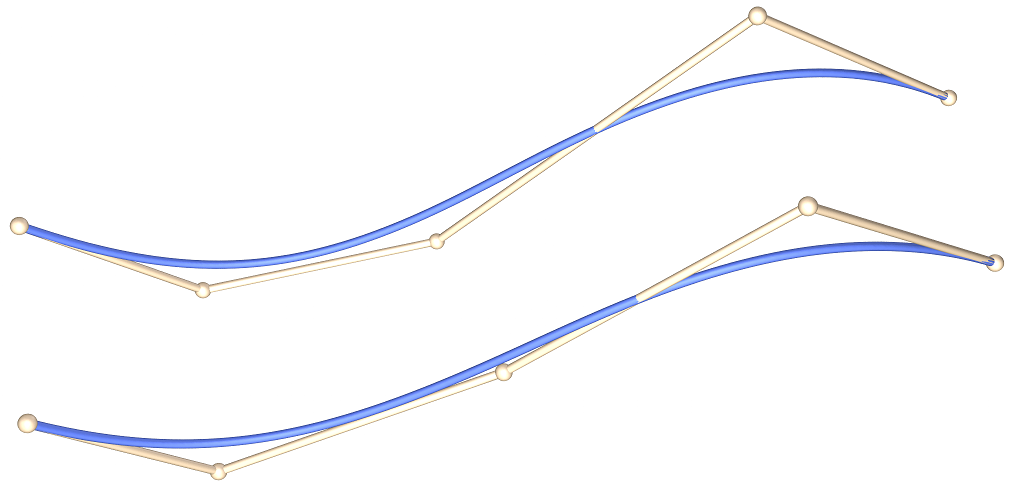}  
}
\hfill
\subfigure[]{
\centering     
\includegraphics[width=0.3\textwidth]{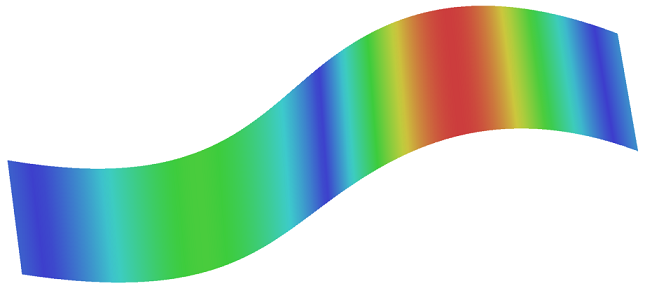}%
}
\hfill
\subfigure[]{
\centering     
\includegraphics[width=0.3\textwidth]{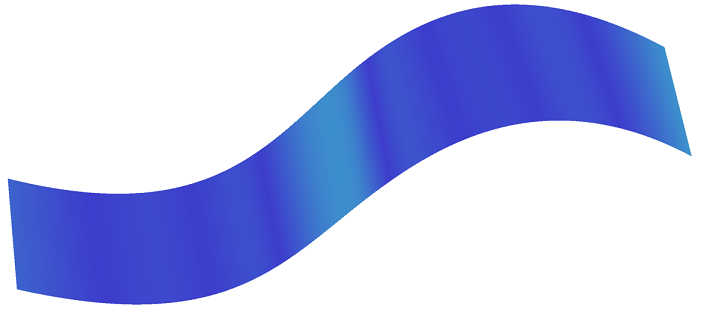}
}}
\end{minipage}
\vfill
\begin{minipage}[b]{0.9\linewidth}
\centering
\centerline{
\hfill
\subfigure[]{
\centering     
\includegraphics[width=0.3\textwidth]{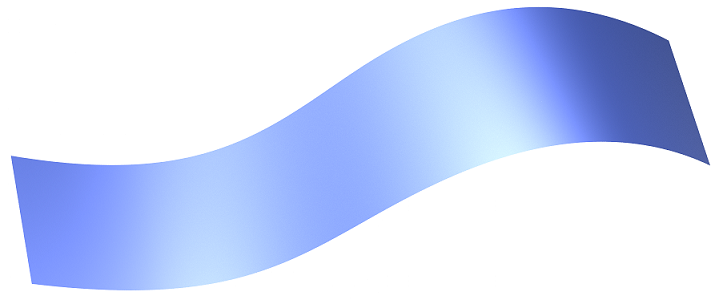}
}\hfill
\subfigure[]{
\centering     
\includegraphics[width=0.3\textwidth]{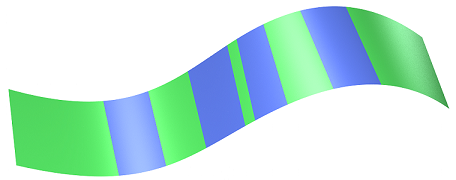}
}\hfill
\subfigure[]{
\centering     
\includegraphics[width=0.3\textwidth]{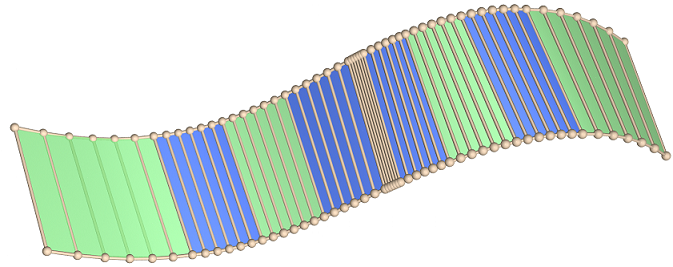}
}\hfill
}
\end{minipage}
\caption{Conversion into a B-spline surface. (a) Input B-spline curves with control points. (b) Initial ruled surface by the simple method. The (maximum,average) warp angle is $(9.27\degree,3.97\degree)$. (c) Result by our method. The (maximum,average) warp angle is $(1.38\degree,0.37\degree)$. (d) Rendered resulting surface. (e) B$\acute{e}$zier patches of the resulting surface. (f) A B-spline representation shown with its control points of the resulting surface. }
\label{fig:bsplineform}
\end{figure}

An example is given in Figure ~\ref{fig:bsplineform} to illustrate this algorithm. 
The input curves are  both degree 3 B-splines curves with 5 control points and the same knot vector   $[0,0,0,0,0.5,1,1,1,1]$. The used  mapping function is a degree 2 B-spline function with 10 control points and a knot vector
$$ [0,0,0,\frac{1}{6}, \frac{5}{18}, \frac{7}{18}, 0.5, \frac{11}{18}, \frac{13}{18}, \frac{5}{6},  1,1,1].$$
The result is a B-spline surface composed of 9 pieces of B$\acute{e}$zier surfaces, whose knot vector for the parameter $t$ is 
$$ [0^{(7)},\frac{1}{9}^{(6)},\frac{2}{9}^{(6)}, \frac{1}{3}^{(6)}, \frac{4}{9}^{(6)}, \frac{5}{9}^{(6)}, \frac{2}{3}^{(6)}, \frac{7}{9}^{(6)}, \frac{8}{9}^{(6)}, 1^{(7)}],  $$ 
where $t^{(p)}$ means a knot value $t$ with a duplication of $p$. We use a clamped uniform  knot vector here which does not change parametrization mapping in each B$\acute{e}$zier surface patch.

When the input  curves $C_1(t)$, $C_2(T)$ have a different number of control points, different degrees or different knot vectors, they need to be  preprocessed to satisfy our assumption. The input curves with different degrees can be transformed to curves with  an identical degree  using the standard degree elevation algorithm ~\cite{PiegTill96TheNurbsBook}.  Once the input curves have the same degree,  different number of control points or different knot vectors of input curves can be resolved by employing the standard knot insertion algorithm.

\section{Experiments and discussions}

In this section, we demonstrate the performance of the proposed method by some examples. The presented algorithm has been implemented with C++ and experiments are conducted on a laptop with  a AMD 2500U CPU and 8GB DDR4 memory.  When the input curves are B-spline curves as the ship-hull models in Figure ~\ref{fig:Ship-hull}, our method is applied directly. When input models are meshes as the examples in Figures ~\ref{fig:felt}, ~\ref{fig:Hatstrip} and ~\ref{fig:shoe},  polylines serving as  boundaries of developable strips are firstly extracted from the model which are then fitted with B-spine curves using traditional curve fitting methods.  

We measure developability of a resulting surface by the warp angles on some sample ruling lines on a surface.  A warp angle $\beta_i$  at a ruling line is  a fabrication related measurement which is defined to be the angle between the surface normals $N(0,t_i)$  and $N(1,t_i)$  at  the endpoints of the ruling line. We evaluate both the maximum warp angle $\beta_{max}=\max\{\beta_i, i=0,...,K\}$ and the average warp angle $\beta_{ave}= \frac{1}{1+K}{\sum_{i=0}^{K}{\beta_i}}$  by taking $K+1$ sample ruling lines on a surface.  To illustrate warp angle variation on a surface, the warp angles $\beta_i$, i=0,...,K are mapped to a color bar and a surface is rendered with the mapped color, with the red color indicating the maximum value and the blue color indicating the smallest value. The set of sample ruling lines for rendering color coding is the same as the samplings in optimization,  except for several surfaces in Figure ~\ref{fig:Helical} which are pointed out explicitly.

\subsection{Influence of degrees and the number of control coefficients in a B-spline mapping function}
Our algorithm uses a B-spline function to approximate an optimal parametrization mapping function, therefore, the approximation ability of the employed function has an essential impact on surface developability.  We experiment with several mapping functions with various degrees including linear, quadratic and cubic functions as well as a various  number of control coefficients in a function.  On one side, we prefer to use a low degree because the final B-spline surface is degree $pd$ and a high degree $d$ of a  mapping function will lead to a B-spline surface with a high degree.  On the other side, a mapping function is required to possess strong enough approximation ability which cannot be provided by a piecewise linear function. 

\begin{figure}[t]
\centering
\begin{minipage}[b]{0.932\linewidth}
\centerline{
\hfill
\subfigure[]{
\begin{overpic}[width=0.15\textwidth]{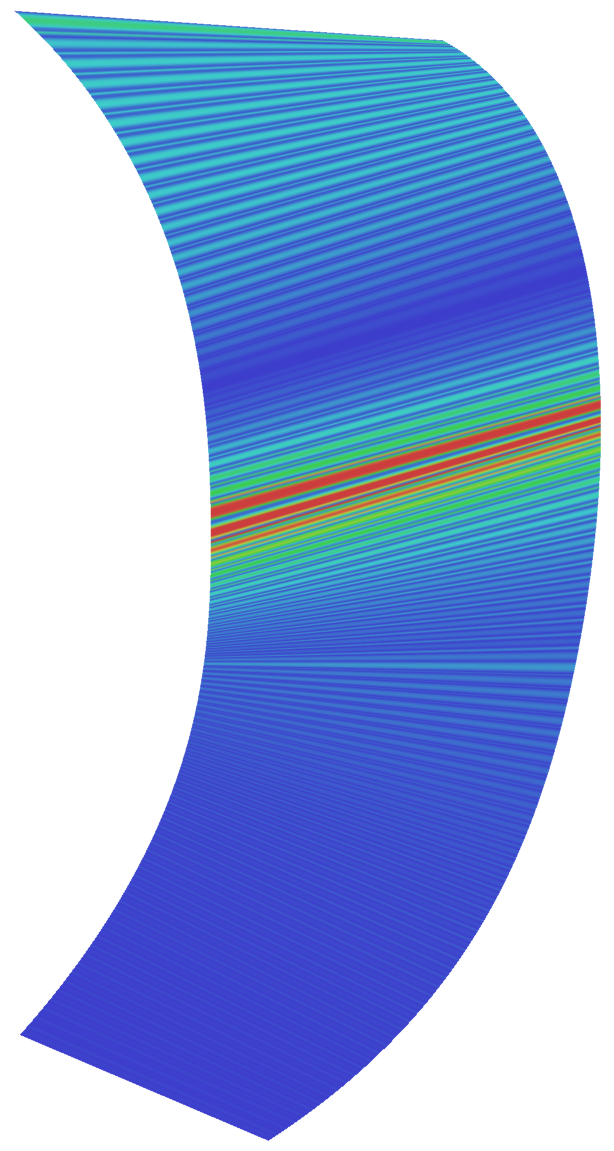}
    	{\small
    		\put(26,5){\fcolorbox{black}{white}{$(1,100)$}}
    	}
 \end{overpic}
 }
\hfill
\subfigure[]{
\begin{overpic}[width=0.15\textwidth]{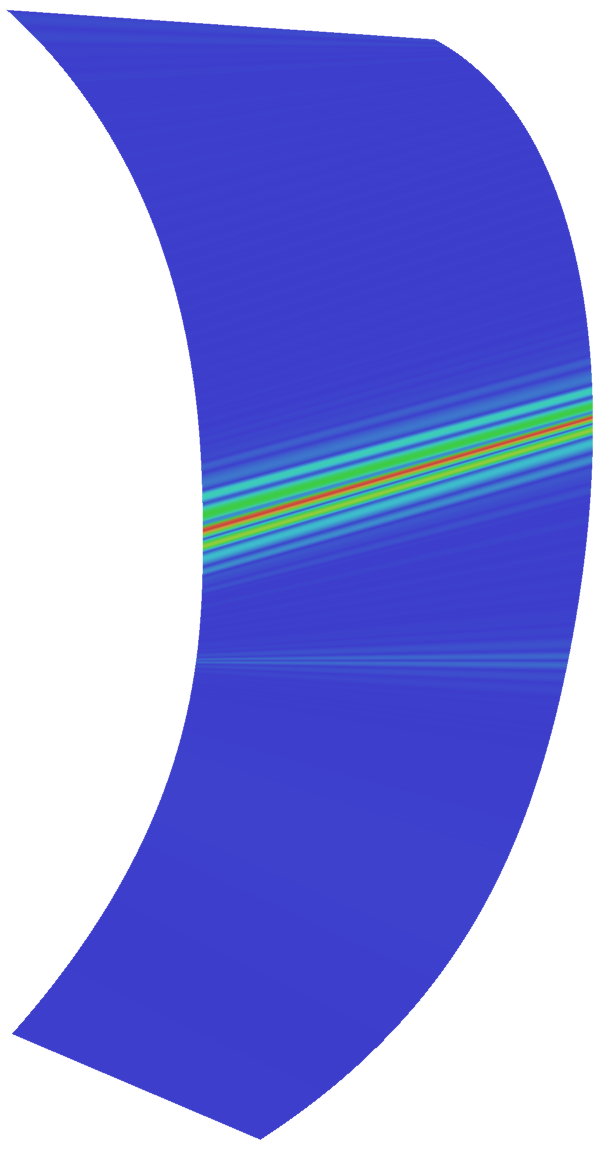}
    	{\small
    		\put(26,5){\fcolorbox{black}{white}{$(2,100)$}}
    	}
 \end{overpic}
 }
 \hfill
 \subfigure[]{
\begin{overpic}[width=0.15\textwidth]{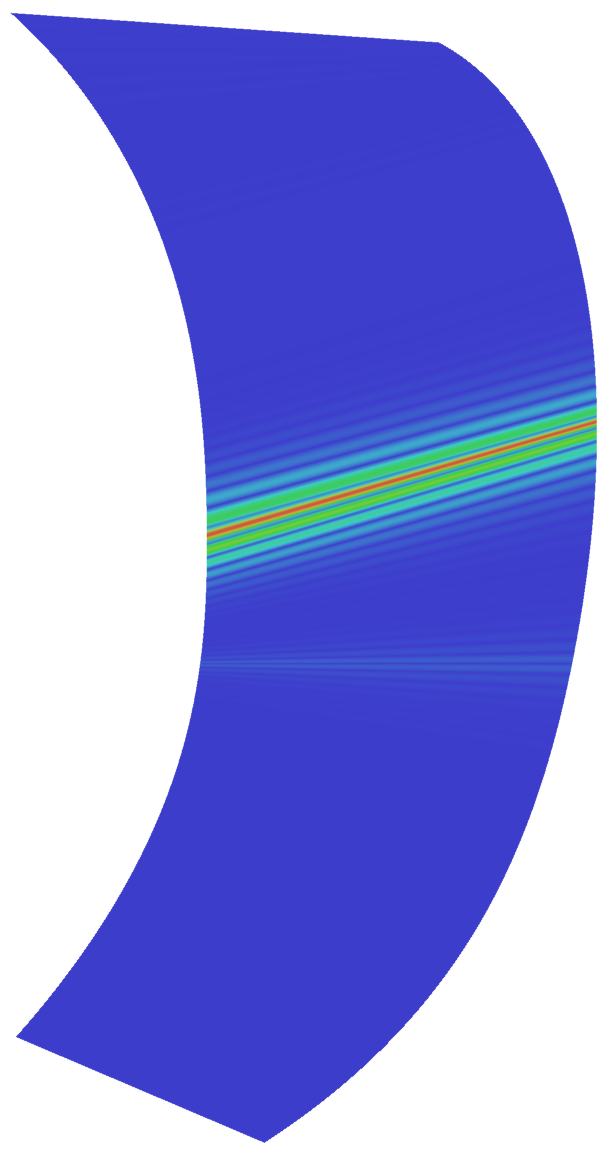}
    	{\small
    		\put(25,5){\fcolorbox{black}{white}{$(3,100)$}}
    	}
 \end{overpic}
 }
 \hfill
 \subfigure[]{
 \begin{overpic}[width=0.15\textwidth]{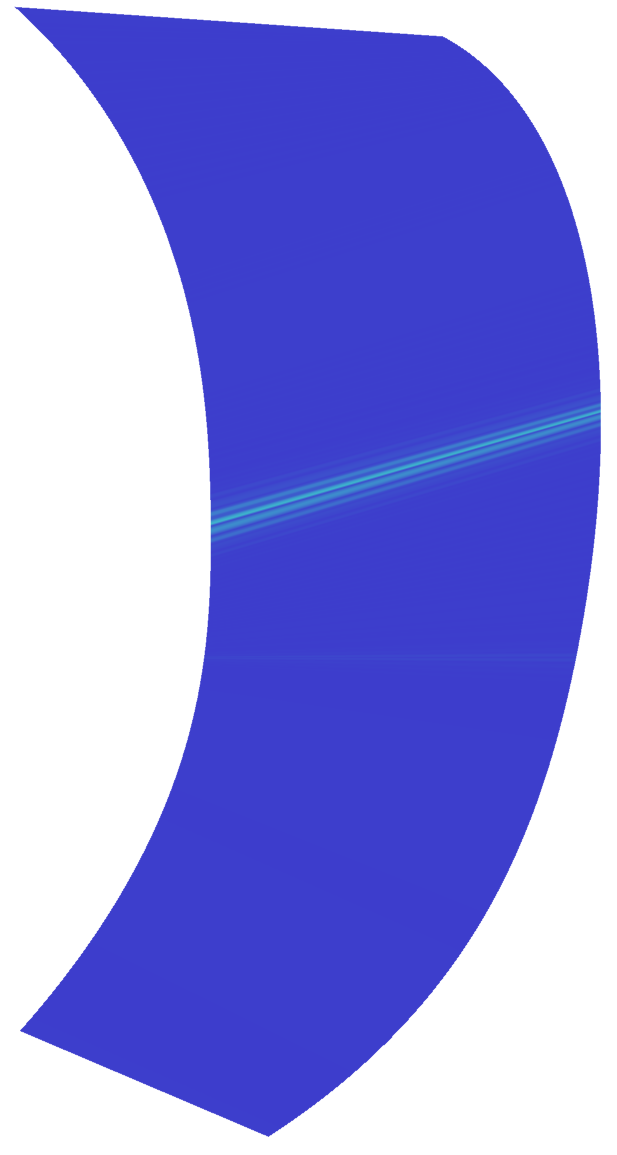}
    	{\small
    		\put(25,5){\fcolorbox{black}{white}{$(2,200)$}}
    	}
 \end{overpic}
 }
 \hfill
\subfigure[]{
\centering     
\includegraphics[width=0.3\textwidth]{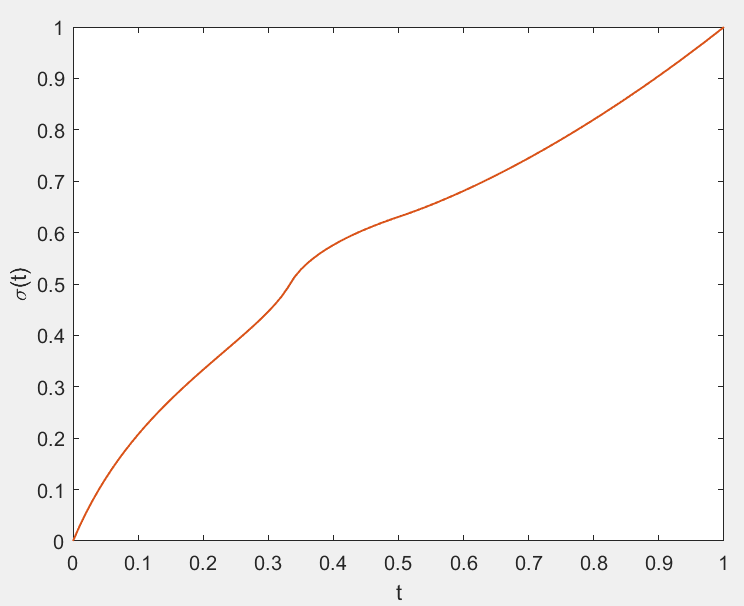}
}
}
\end{minipage}
\vfill
\begin{minipage}[b]{0.95\linewidth}
\centering
\centerline{
\hfill
\subfigure[]{
\centering     
\includegraphics[width=0.3\textwidth]{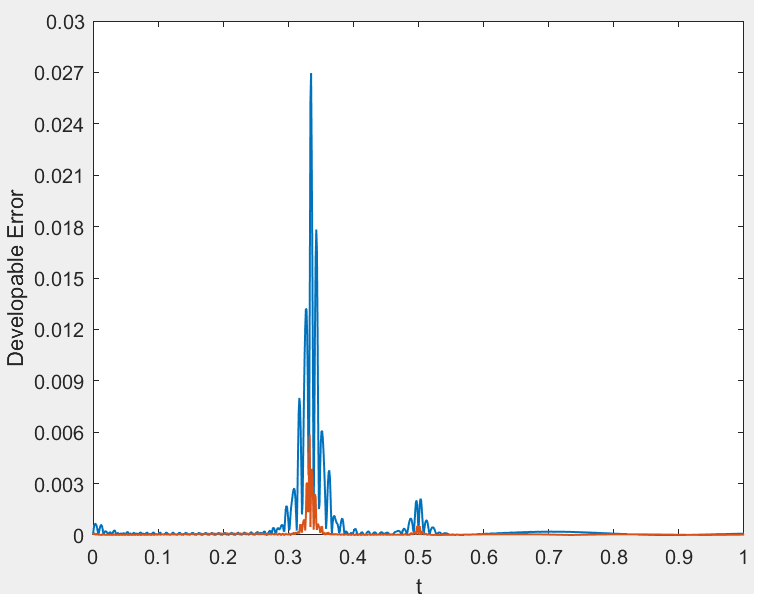}
}
\hfill
\subfigure[]{
\centering     
\includegraphics[width=0.29\textwidth]{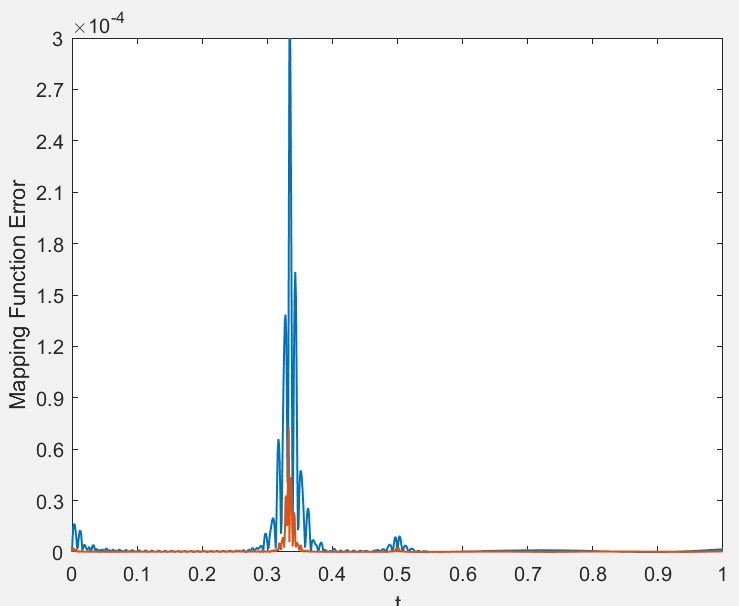}
}
\hfill
\subfigure[]{
\centering     
\includegraphics[width=0.3\textwidth]{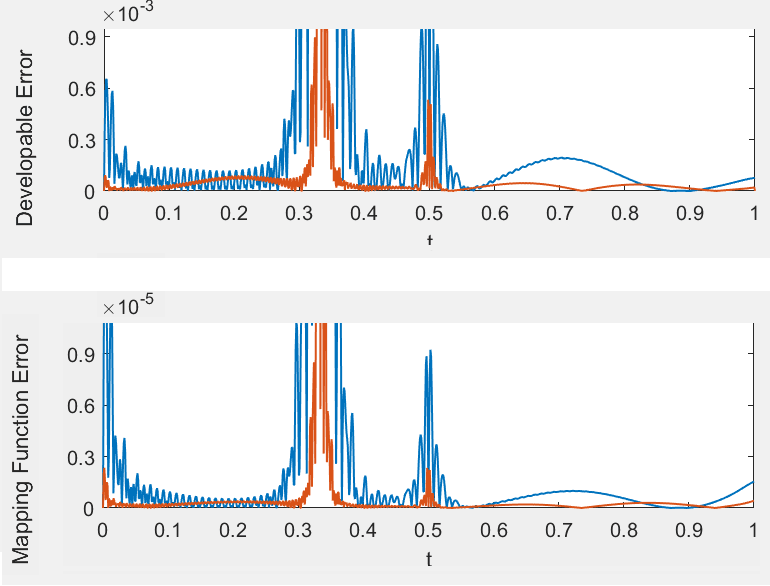}
}\hfill
}
\end{minipage}
\caption{Experiments with varying degree and  a number of control coefficients in B-spline mapping functions. (a)-(c) The results using various degrees with the same number of control coefficients. The small boxes besides each figure means $(\textit{degree},\textit{\#control coefficients})$. The (maximum,average) warp angle for the surfaces in (a)-(d) is respectively (0.04\degree,0.0026\degree),  (0.026\degree,$5.5\cdot 10^{-4}$\degree),  (0.026\degree,$5.2\cdot 10^{-4}$\degree) and (0.0058\degree,$8.2\cdot 10^{-5}$\degree). The range of warp angles [0$\degree$, 0.027$\degree$] is used for color mapping in all figures. (e) The accurate parametrization mapping function. (f)-(g) Surface developability and approximation errors using a quadratic B-spline mapping function using 100 and 200 control coefficients respectively. The red line refers to the one with 200 control points. (h) Zoom in windows of the graphs in (f) and (g).  }
\label{fig:DifferentDegree}
\end{figure}

In Figure~\ref{fig:DifferentDegree}, we show the results using various degrees and numbers of control coefficients in a B-spline mapping  function.  The input curves  in this example define an exact developable surface with a known parametrization mapping function shown in Figure ~\ref{fig:DifferentDegree}(e). Figures ~\ref{fig:DifferentDegree}(a)-(c) show the results using B-spline functions of  degrees 1, 2,3 respectively, all are defined with 100 control coefficients.  We observe that using a quadratic B-spline function can achieve much better developability than using a  linear B-spline function when the same number of control coefficients are used. However, increasing the  function  degree  further to degree 3 does not produce much improvement on surface developability. Figure~\ref{fig:DifferentDegree} (d) is a result using a quadratic B-spline function with 200 control coefficients. It shows that  increasing the number of control coefficients from 100 to 200 greatly improves surface developability.  

Figures ~\ref{fig:DifferentDegree}(f),(g) illustrate surface developability and approximation errors of the optimized mapping functions, measured at some sample ruling lines,  corresponding to the surfaces in Figures~\ref{fig:DifferentDegree}(b), (d). This demonstrates the performance of our algorithm when different number of control coefficients are used in a B-spline mapping function. The approximation errors of a mapping function $\sigma(t)$ is defined by $\epsilon(t)=\| \sigma(t)-\sigma^*(t) \|$ where $\sigma^*(t)$ denotes the accurate parametrization mapping function. We observe that using more control points in a B-spline function provides a stronger approximation ability and consequently a higher order of developability of the resulting surface can be obtained. Another interesting phenomenon  we observe  is a similar distribution of surface developability  and function approximation errors illustrated by Figures  ~\ref{fig:DifferentDegree}(f)-(h).

\subsection{Parametrization mapping function and surface developability}
When existing methods for polyline boundary curves are adapted to  smooth input curves, the sequence  of lines  with large developability is found in a finite set of lines connecting sample points on  input curves.  The lack of  a capability of continuous searching on input curves makes the existing methods unable to find a ruling line achieving the largest developability in a local neighborhood. Although the develoapability of the lines can be fairly good when the a dense sampling is employed, we show that, in fact, a tiny difference in a mapping function can lead to a considerable  difference in surface developability.  

\begin{figure}[t]
\centering
\begin{minipage}[b]{0.9\linewidth}
\centerline{
\hfill
\subfigure[]{
\centering     
\includegraphics[width=0.28\textwidth]{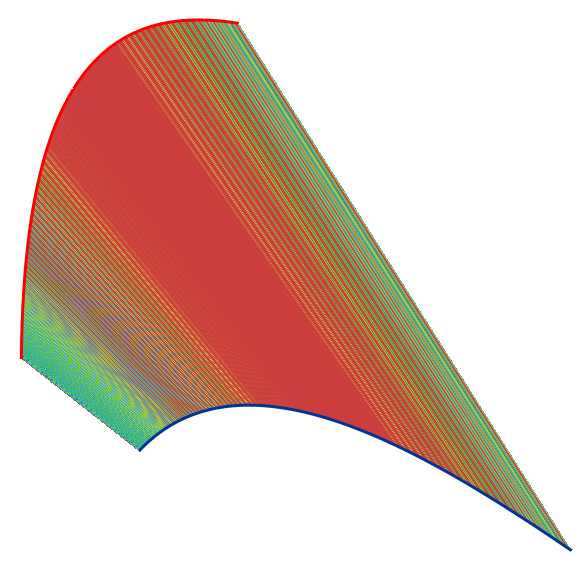}
}
\hfill
\subfigure[]{
\centering     
\includegraphics[width=0.3\textwidth]{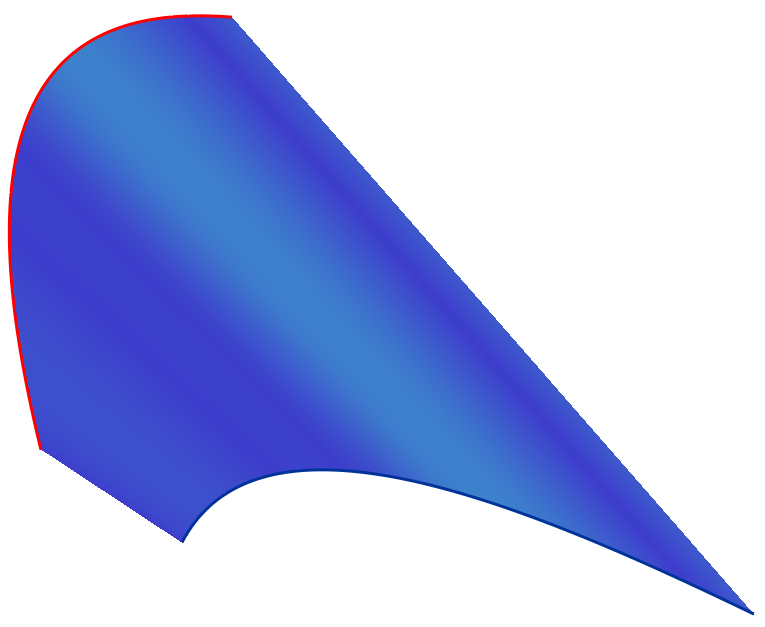} 
}
\hfill
\subfigure[]{
\centering     
\includegraphics[width=0.4\textwidth]{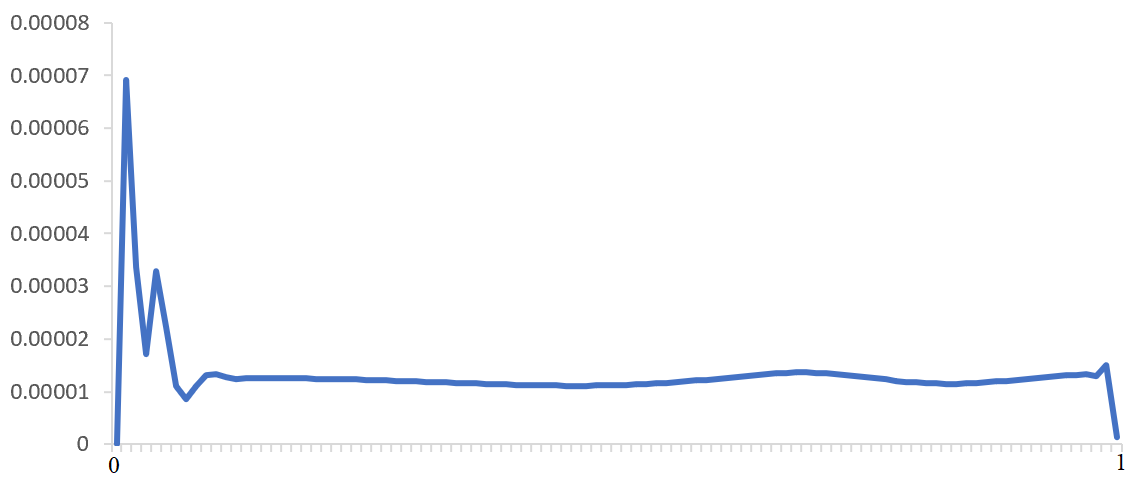}
}\hfill
}
\end{minipage}
\caption{The impact of a mapping function on  surface developability.  (a) The result using a linear B-spline function with 100 shape parameters. The (max,average) warp angle is
 ($3\cdot 10^{-3}$\degree,$7\cdot 10^{-4}$ \degree).   (b) The result using a quadratic B-spline function with 3 control coefficients. The (max,average) warp angle is $(6\cdot 10^{-5}$\degree, $1\cdot 10^{-5}$\degree). The range of warp angles for color coding is 
 [0\degree, $5\cdot 10^{-4}$\degree] for both images. (c) shows the difference of mapping functions in (a) and (b). }
\label{fig:degree2}
\end{figure}

This is demonstrated by an example in Figure ~\ref{fig:degree2} where both degree 1 and degree 2 B-spline mapping functions are tested with a simple model.   There exists  an exact developable surface  with the input curves with a mapping function  $T=t^2$.  The linear B-spline function we use  has 100 control coefficients  and the quadratic B-spline function has only 3 control coefficients.   Figure ~\ref{fig:degree2}(c) shows  a graph of the difference, defined by $\| \sigma_1(t)-\sigma_2(t) \|$,   between  a linear B-spline function $\sigma_1(t)$ and a quadratic B-spline function $\sigma_2(t)$, both obtained with  our optimization algorithm.  We observe that the maximum  difference between these two functions is around $7 \cdot 10^{-5}$. The average warp angle of the surface defined by $\sigma_2(t)$ is only $1/70$ of the value by $\sigma_1(t)$. A tiny difference in the optimized parametrization mapping function produces a considerable improvement on surface developability, which means a huge number of samplings with existing methods. Therefore, this experiment is also a demonstration of the importance of a continuous searching ability on input curves.

\subsection{Comparison with the discrete mapping method}

In order to obtain a smooth ruled surface, exisiting methods create a smooth surface by computing smooth boundary curves interpolating   the end  points of a sequence of ruling lines ~\cite{Perez2007}.  One of the drawbacks of this kind of methods is that the final smooth surface depends on the parametrization of the data points in curve interpolation.  

\begin{figure}[t]
\centering
\begin{minipage}[b]{1\linewidth}
\subfigure[]{
\centering     
\includegraphics[width=0.1\textwidth]{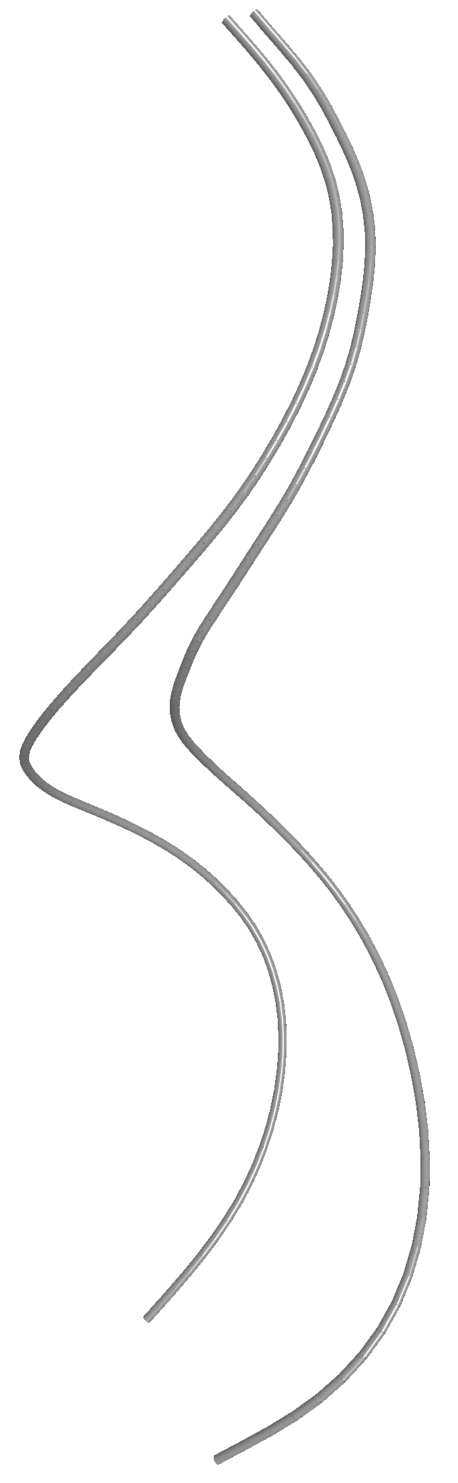} 
}
\subfigure[]{
\centering     
\includegraphics[width=0.1\textwidth]{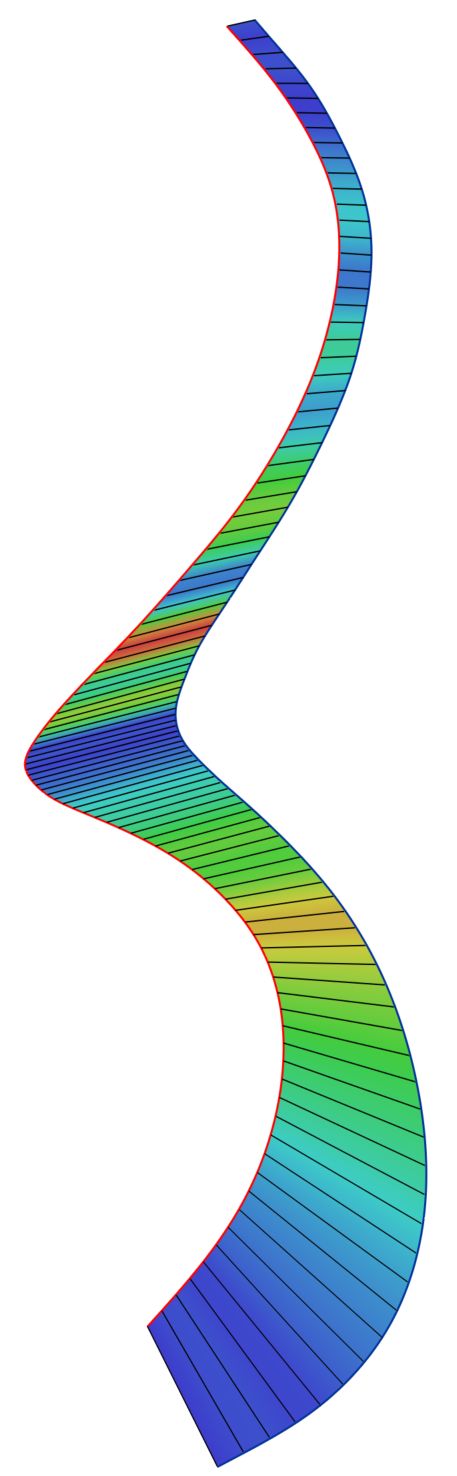}
}
\subfigure[]{
\centering     
\includegraphics[width=0.1\textwidth]{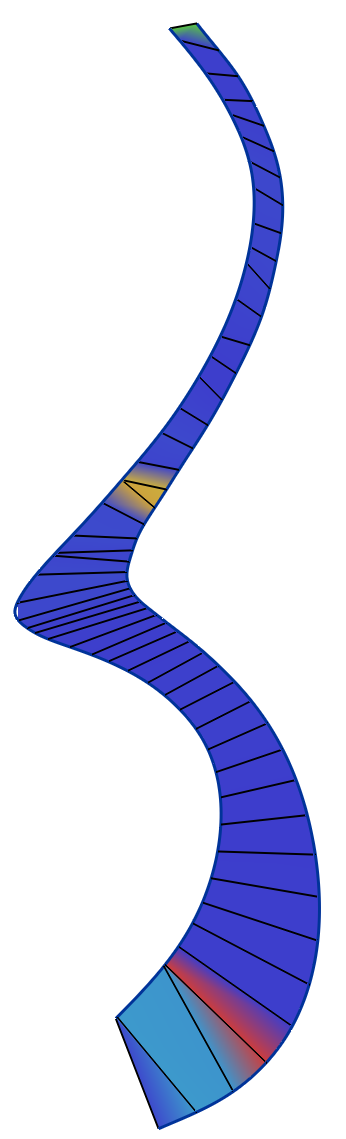} 
}
\subfigure[]{
\centering     
\includegraphics[width=0.094\textwidth]{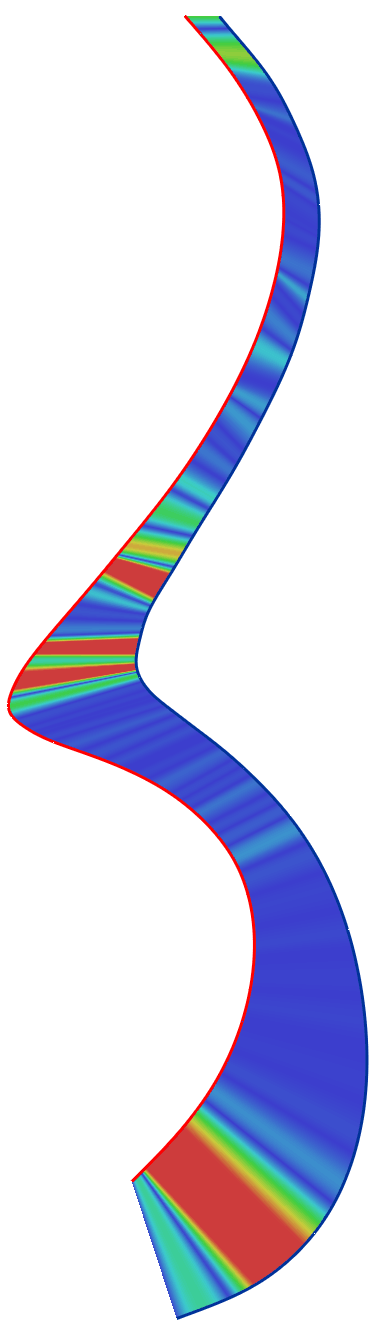}
}
\subfigure[]{
\centering     
\includegraphics[width=0.094\textwidth]{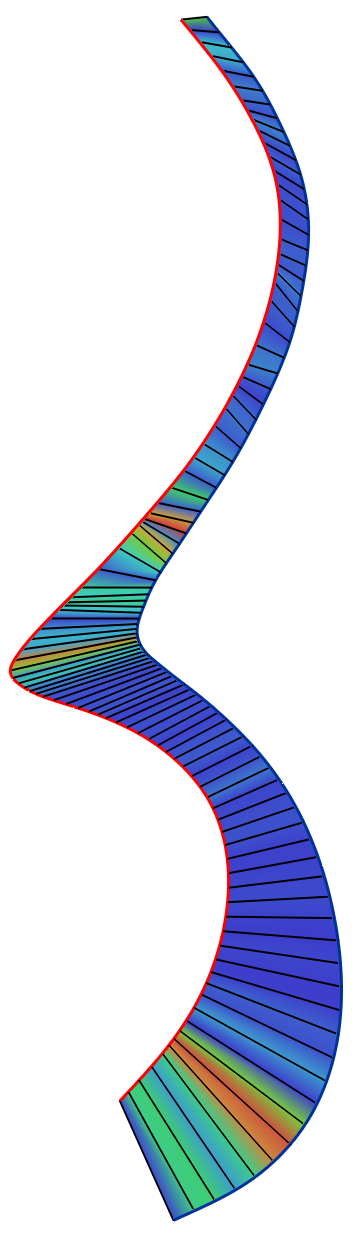} 
}
\subfigure[]{
\centering     
\includegraphics[width=0.095\textwidth]{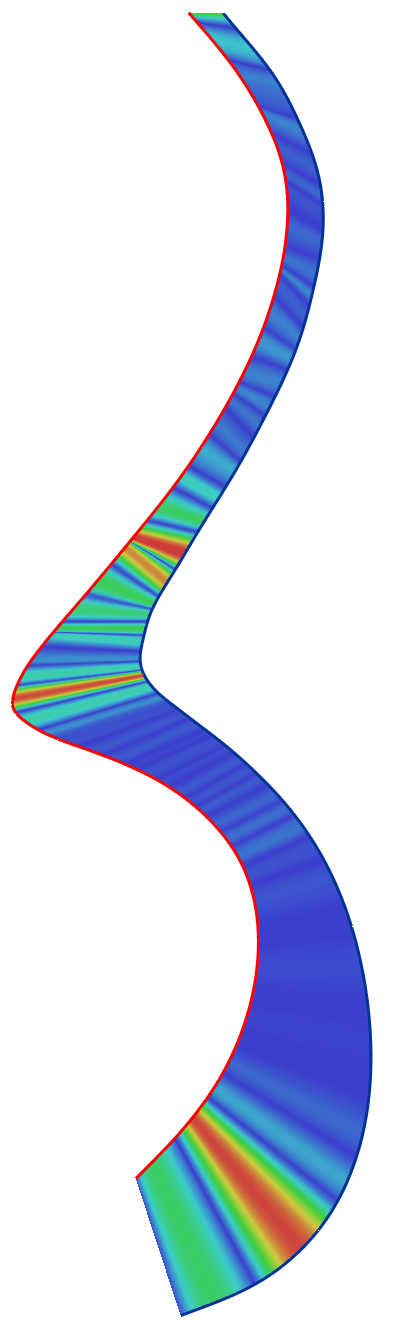}
}
\subfigure[]{
\centering     
\includegraphics[width=0.288\textwidth]{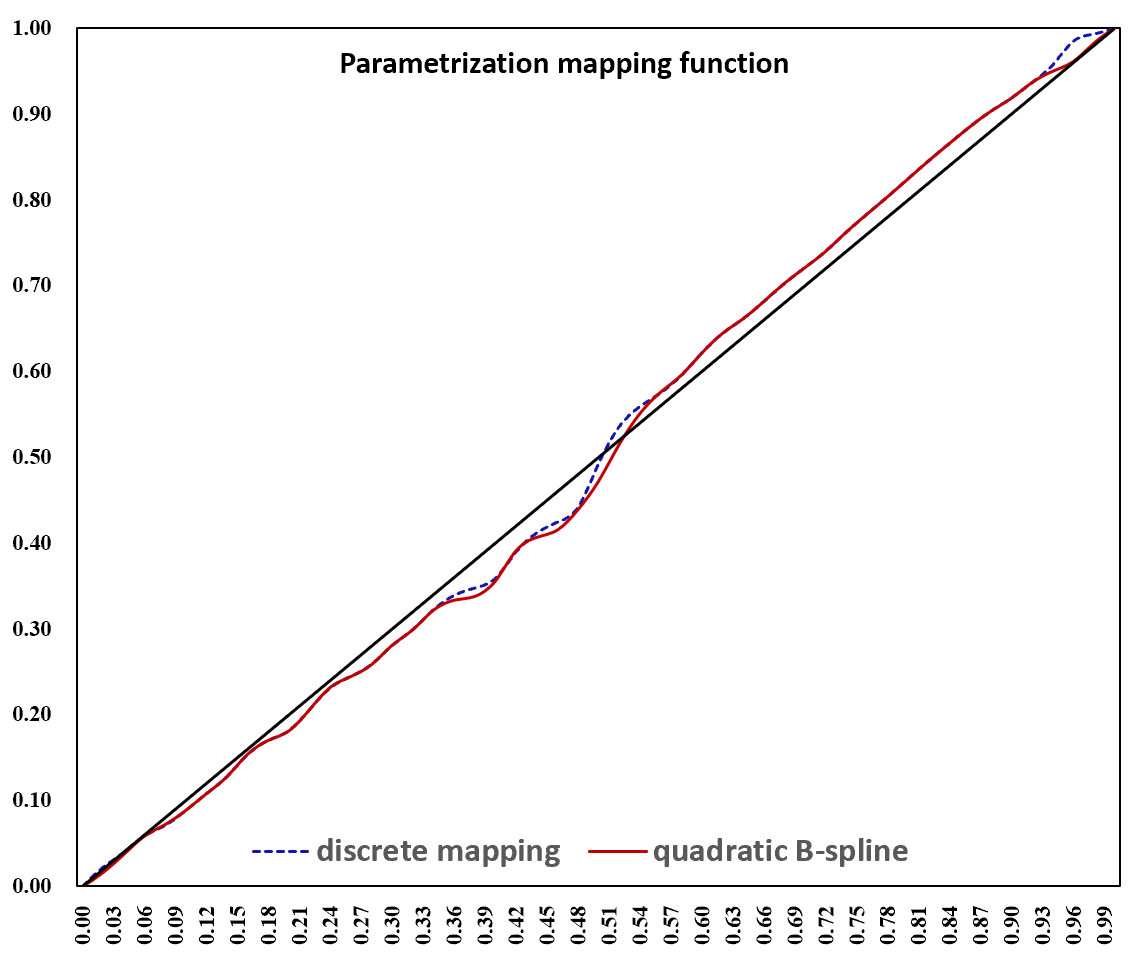}
}\hfill
\end{minipage}
\caption{Comparison between a discrete mapping function with a quadratic mapping function. (a) Input curves. (b) A ruling surface defined by a simple mapping function $T=t$, which  acts an initialization for optimization. The (maximum,average) warp angle is (15.7$\degree$,4.9$\degree$), which are mapped to a color range from red to blue.  (c)  A result using the discrete mapping method. The (maximum,average) warp angle is (1.06$\degree$,0.08$\degree$).  (d) A continuous surface defined  by a continuous mapping function interpolating the results in (c).  The (maximum,average) warp angle is (2.1$\degree$,0.25$\degree$). (e) A result computed using a quadratic B-spline parametrization mapping function. The (maximum,average) warp angle is (1.10$\degree$,0.177$\degree$).  (f) The surface in (e) is rendered by taking 1000 sampling ruling lines. The (maximum,average) warp angle is (1.14$\degree$,0.179$\degree)$. The same range  of warp angles [0$\degree$,1.1$\degree$] is used for the color codings in (c)-(f).  (g) Parametrization mapping functions before and after optimization, where the red curve refers to the optimized result. }
\label{fig:Helical}
\end{figure}

Another idea to adapt existing method to produce a smooth ruled surface is to compute  a continuous mapping function interpolating the parameters $(t_i,T_i)$ corresponding to the ruling lines with large developability.  This kind of methods, however, does not guarantee  the developability of surface regions  between  the specified ruling lines. We show by an example in Figure ~\ref{fig:Helical} that surface regions between the specific ruling lines becomes much worse than the specified ruling lines. In this example, the specific ruling lines are computed using the discrete mapping method provided in Section  ~\ref{sec:variarional_framework} which is a continuous searching method and is thus more powerful than existing methods in finding ruling lines with large developability. 

We use 50 samples in the discrete mapping method and obtain a result  in Figure ~\ref{fig:Helical} (c) in which the colors are assigned to the ruling lines obtained by optimization.  Figure ~\ref{fig:Helical}(d) illustrate a surface defined by a continuous mapping function obtained via function interpolation to the result in (c), where 1000 sampling rulings are used in developability measurement and color rendering. We observe that  surface developability between  the specific ruling lines from optimization can be much worse than the ruling lines.  Increasing the number of samples  in the discrete mapping  can certainly improve the result which however also increases the number of unknown variables in optimization.

In order to provide a fair comparison to the result using a B-spline function, we use the same number of control coeffcients, i.e., 50 control coefficients, in a B-spline mapping function.  In optimization, 100 samples of the mapping function are used for evaluating surface developability. The resulting surface is shown in Figure ~\ref{fig:Helical}(e) in which colors are assigned to the ruling lines corresponding to the samples in optimization.  The same surface is rendered in Figure ~\ref{fig:Helical}(f) again with 1000 sample rulings for color coding, which exhibits  a similar distribution of  warp angles as Figure ~\ref{fig:Helical}(e). Surface developability measured at 1000 sample ruling lines are almost the same as the samplings  from optimization.

\subsection{More examples}

\begin{figure}[tp]
\centering
\begin{minipage}[b]{0.9\linewidth}
\centerline{
\subfigure[]{
\centering     
\includegraphics[width=0.5\textwidth]{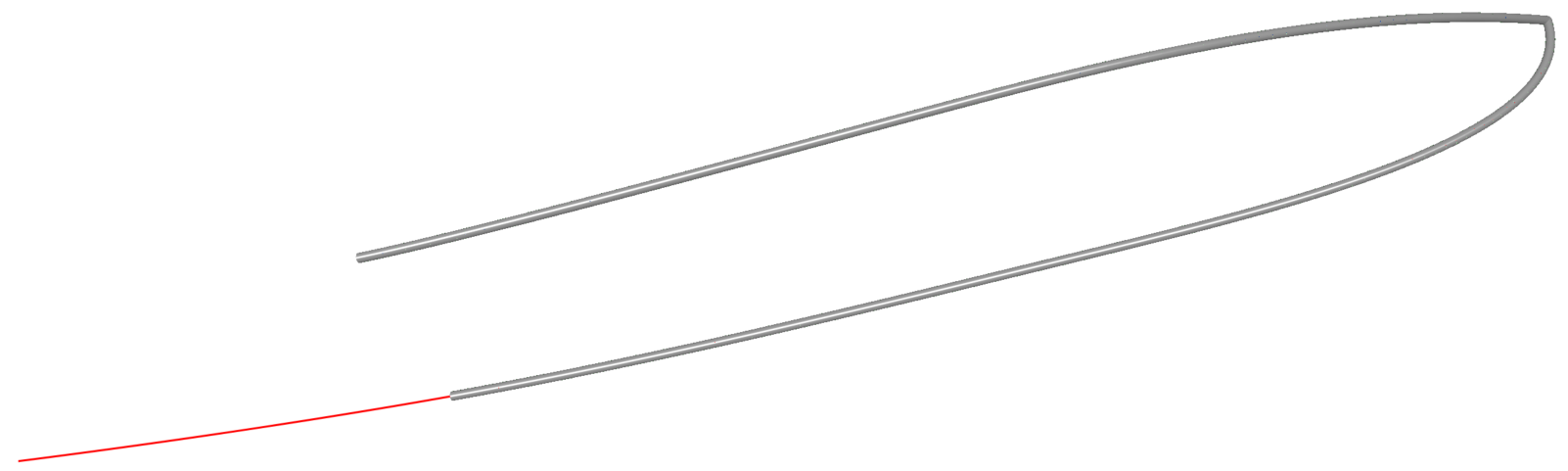}  
}\hfill
\subfigure[]{
\centering     
\includegraphics[width=0.5\textwidth]{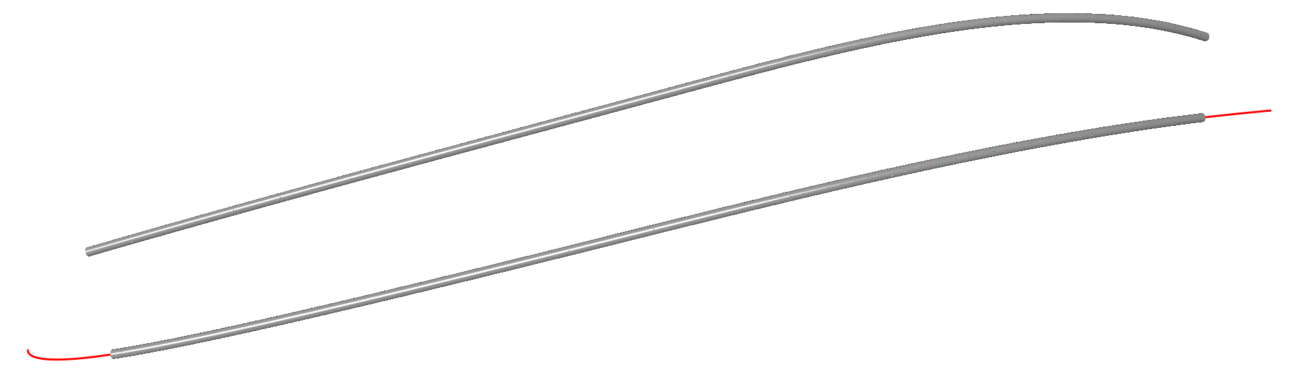}  
}}
\end{minipage}
\begin{minipage}[b]{0.9\linewidth}
\centerline{
\subfigure[]{
\centering     
\includegraphics[width=0.5\textwidth]{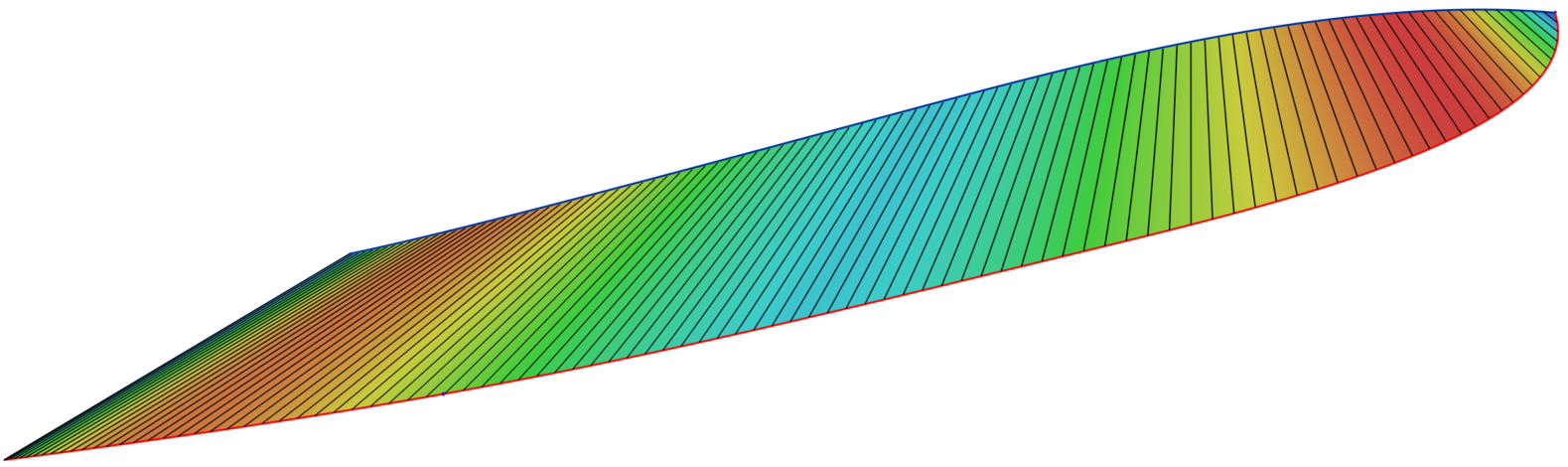}
}\hfill
\subfigure[]{
\centering     
\includegraphics[width=0.5\textwidth]{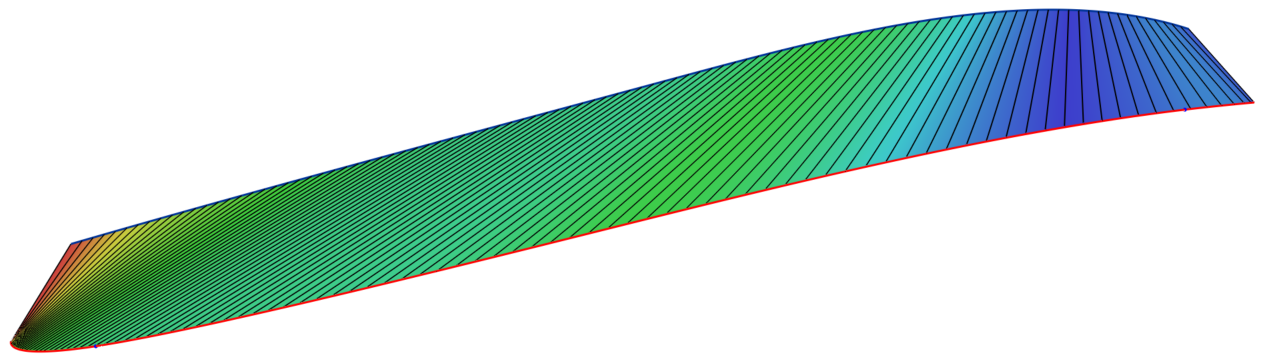}
}}
\end{minipage}
\vfill
\begin{minipage}[b]{0.9\linewidth}
\centerline{
\subfigure[]{
\centering     
\includegraphics[width=0.5\textwidth]{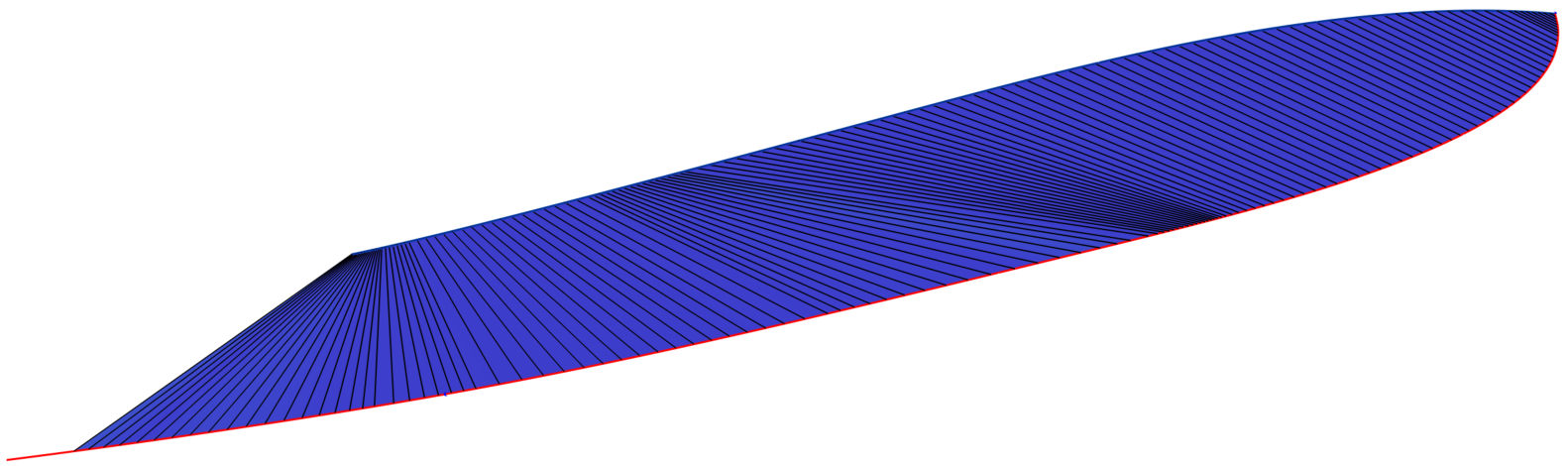}
}\hfill
\subfigure[]{
\centering     
\includegraphics[width=0.5\textwidth]{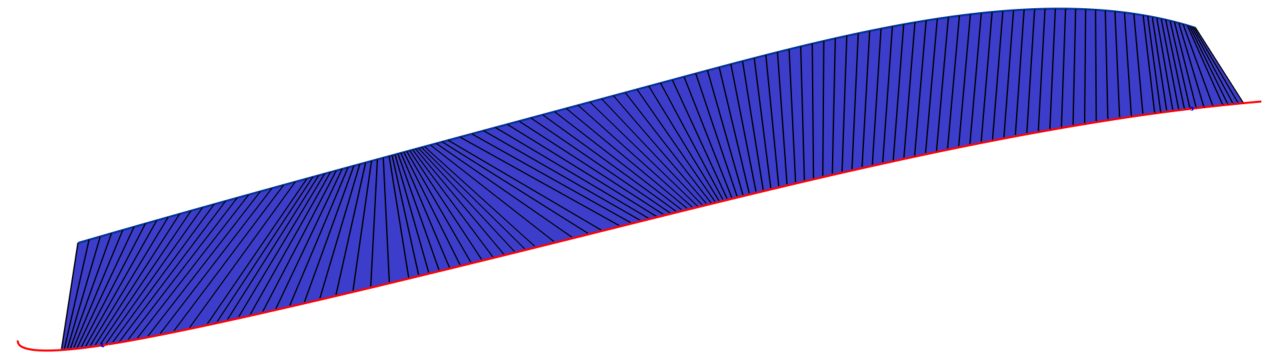}
}
}\end{minipage}
\begin{minipage}[b]{0.9\linewidth}
\centerline{
\subfigure[]{
\centering     
\includegraphics[width=0.45\textwidth]{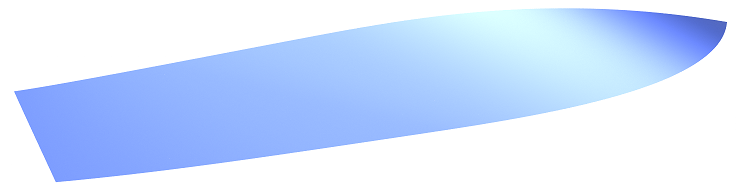}
}\hfill
\subfigure[]{
\centering     
\includegraphics[width=0.45\textwidth]{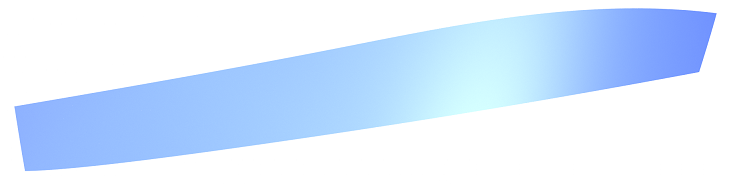}
}
}
\end{minipage}
\begin{minipage}[b]{0.9\linewidth}
\centerline{
\hfill
\subfigure[]{
\centering     
\includegraphics[width=0.3\textwidth]{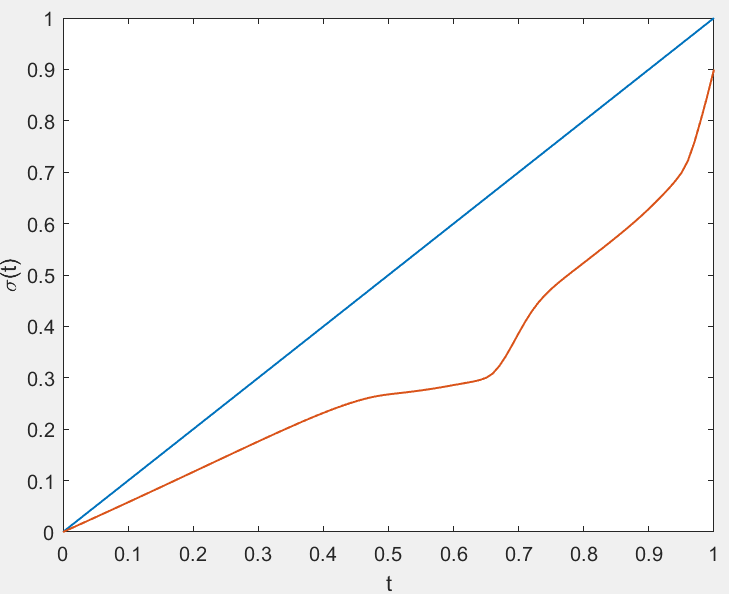}
}\hfill
\subfigure[]{
\centering     
\includegraphics[width=0.3\textwidth]{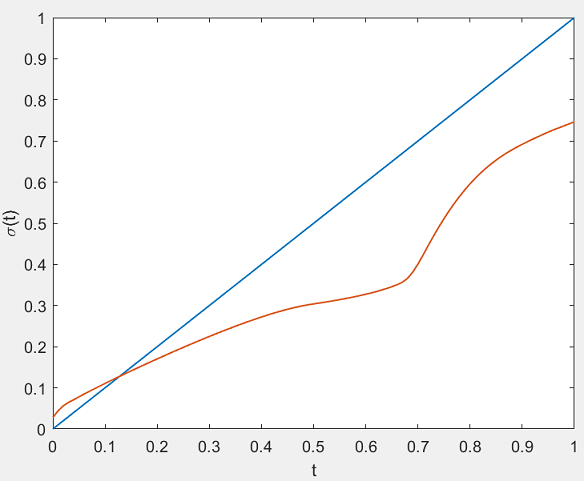}
}\hfill
}
\end{minipage}
\begin{minipage}[b]{0.9\linewidth}
\centerline{
\subfigure[]{
\centering     
\includegraphics[width=0.45\textwidth]{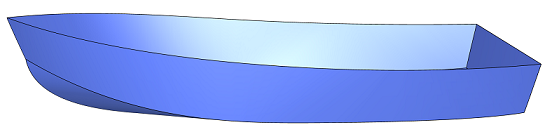}
}\hfill
\subfigure[]{
\centering     
\includegraphics[width=0.45\textwidth]{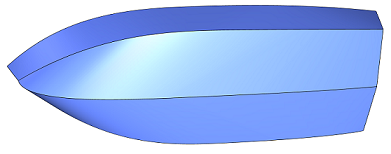}
}
}
\end{minipage}
\caption{A ship-hull model. The left column corresponds to the Chine-Centerline patch and the right column to the Sheer-Chine patch.  (a) and (b) are initial ruled surfaces whose (maximum,average) warp angles are (12.6$\degree$,7.7$\degree$) and  (39.0$\degree$,15.1$\degree$), respectively.  (e) and (f) are the resulting surfaces  whose (maximum,average) warp angles are (0.5$\degree$,0.01$\degree$)  and (0.068$\degree$,0.005$\degree$), respectively. (g)-(h) Rendered strip models.  (i)-(j) Parametrization mapping functions before and after  optimization where a blue curve refers to the optimized function and a red curve refers to the initial function. (k)-(l) Rendered models after trimming. For each strip, warp angles [0$\degree$, maximum degree of the initial ruled surface] is mapped to  a color coding from blue to red.}
\label{fig:Ship-hull}
\end{figure}

\begin{figure}[tp]
\centering
\begin{minipage}[b]{0.9\linewidth}
\centerline{
\subfigure[]{
\centering     
\includegraphics[width=0.27\textwidth]{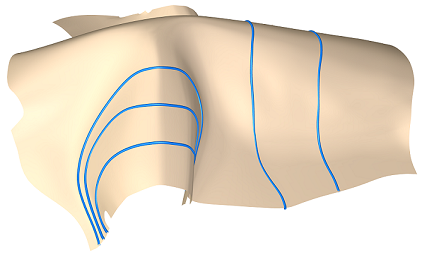}
}
}
\end{minipage}
\begin{minipage}[b]{0.9\linewidth}
\centerline{
\subfigure[]{
\centering     
\includegraphics[width=0.25\textwidth]{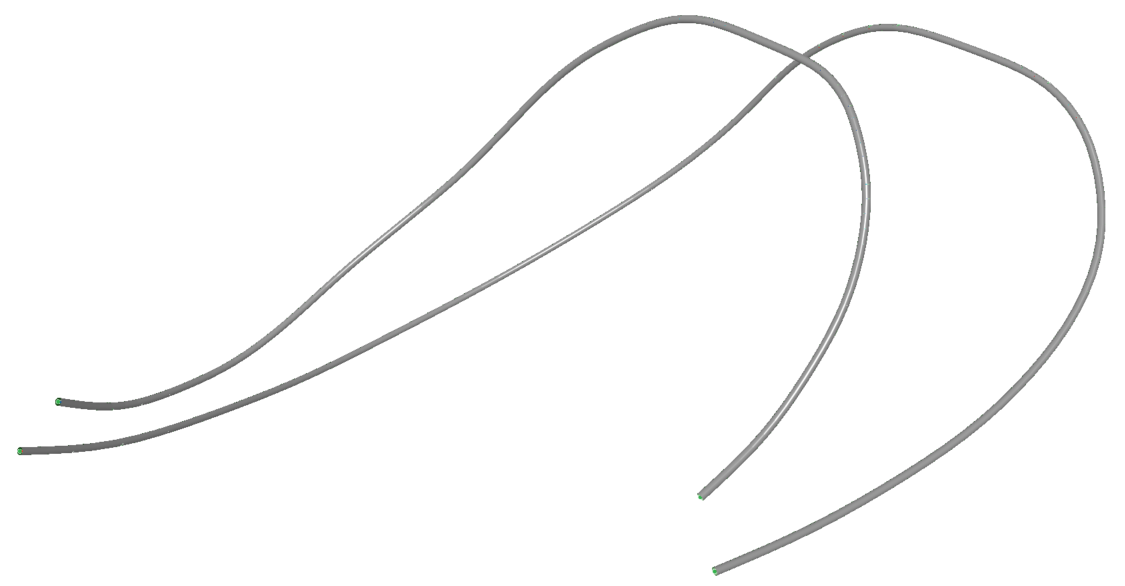} 
}
\hfill
\subfigure[]{
\centering     
\includegraphics[width=0.25\textwidth]{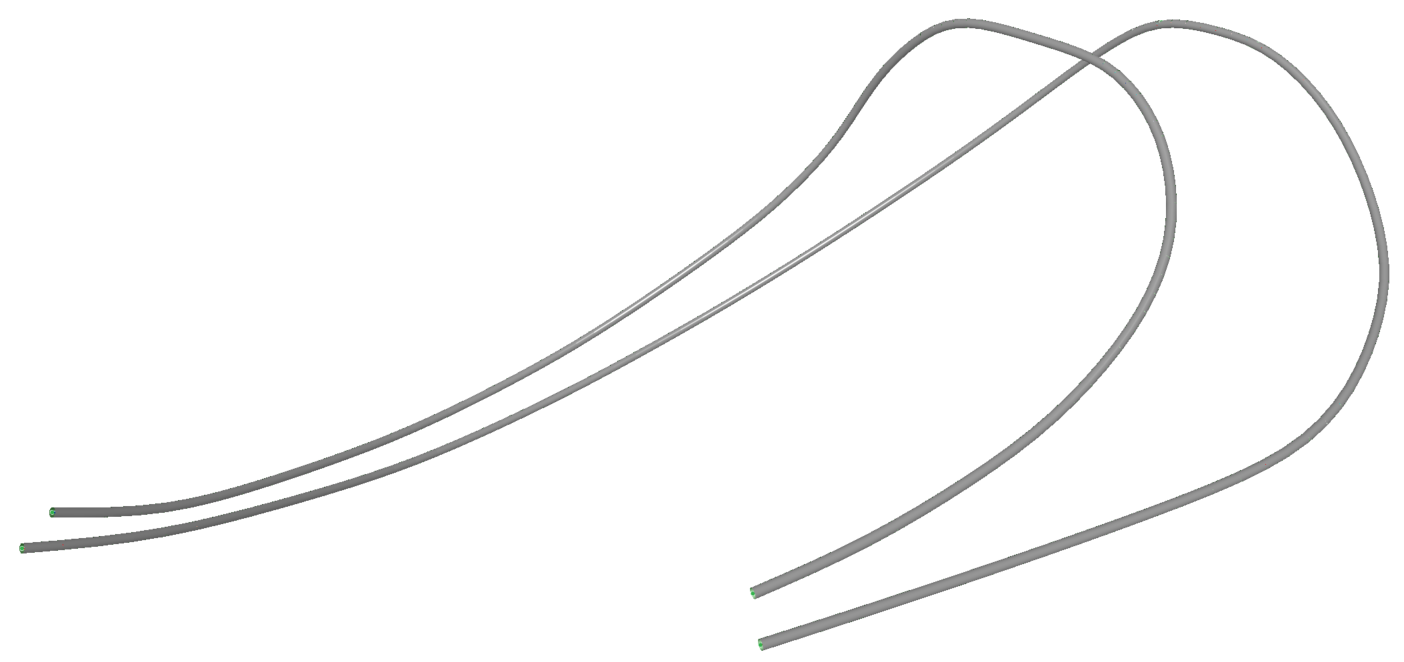} 
}
\hfill
\subfigure[]{
\centering     
\includegraphics[width=0.25\textwidth]{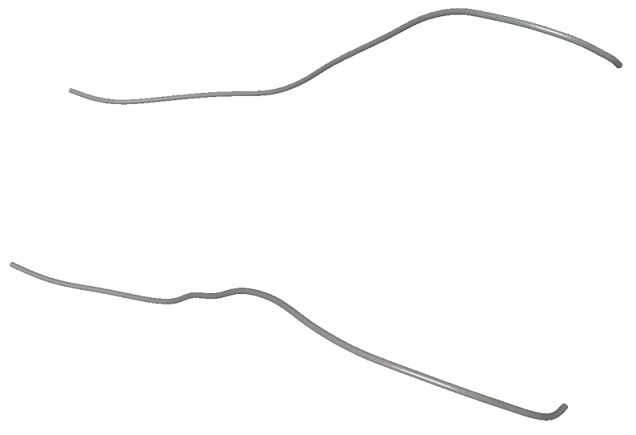}  
}
}
\end{minipage}
\begin{minipage}[b]{0.9\linewidth}
\centerline{
\subfigure[]{
\centering     
\includegraphics[width=0.25\textwidth]{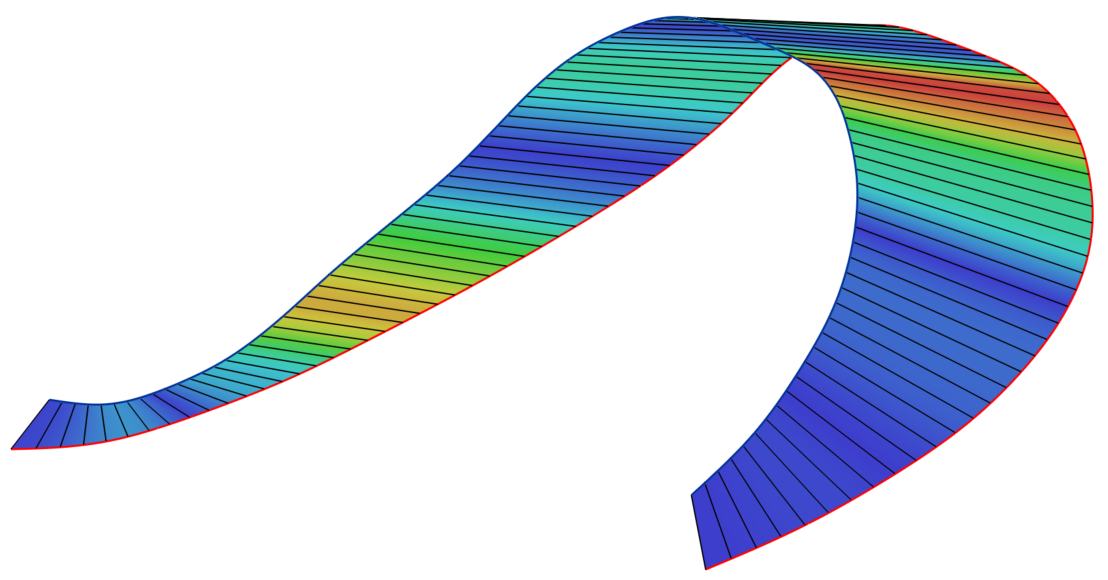}
}
\hfill
\subfigure[]{
\centering     
\includegraphics[width=0.25\textwidth]{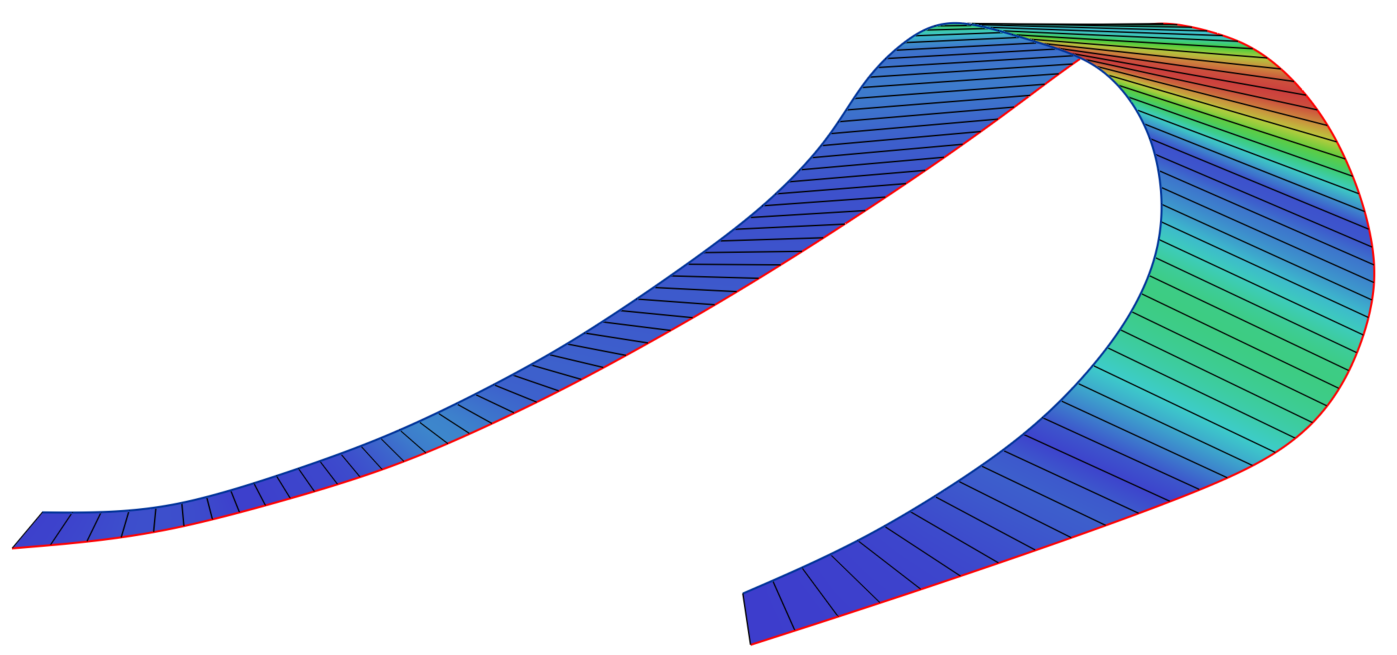}
}
\hfill
\subfigure[]{
\centering     
\includegraphics[width=0.25\textwidth]{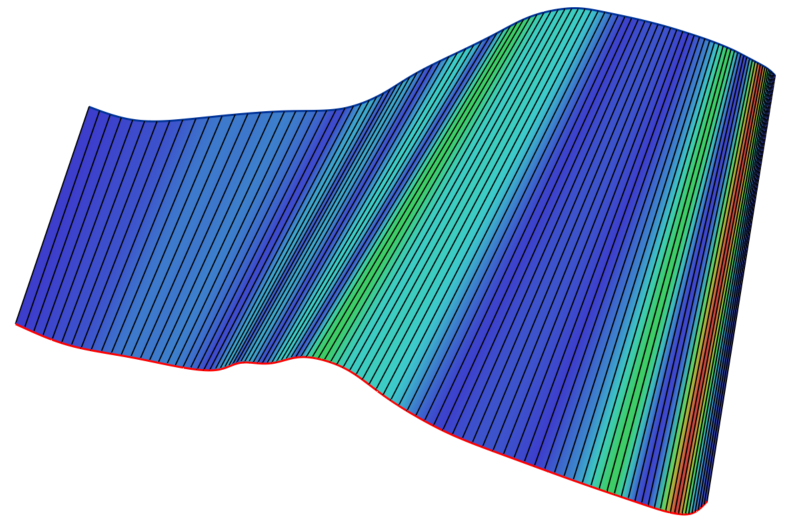}
}
}
\end{minipage}
\begin{minipage}[b]{0.9\linewidth}
\centerline{
\subfigure[]{
\centering     
\includegraphics[width=0.25\textwidth]{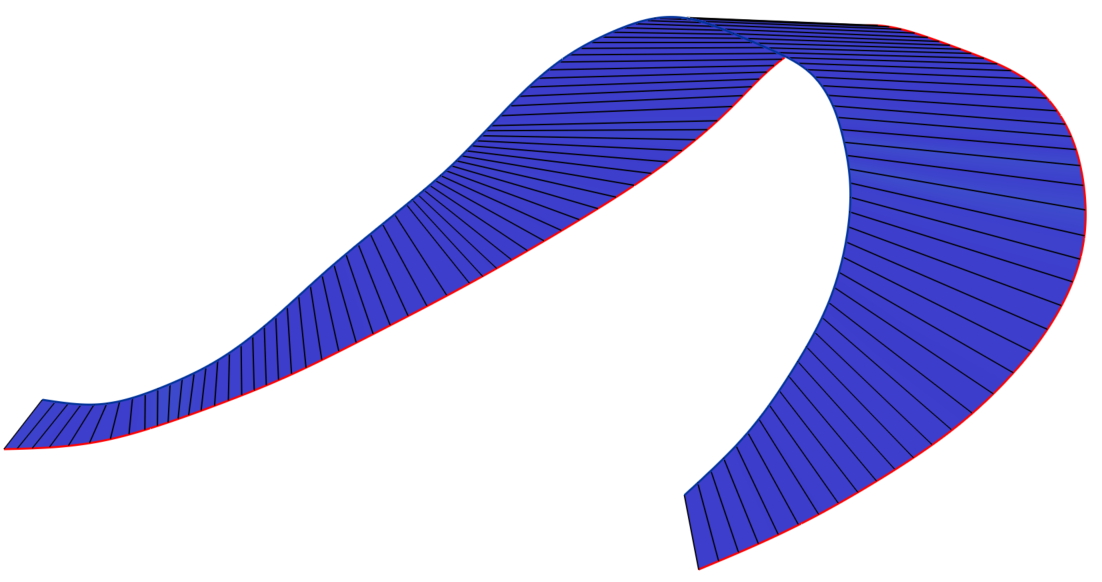}
}
\hfill
\subfigure[]{
\centering     
\includegraphics[width=0.25\textwidth]{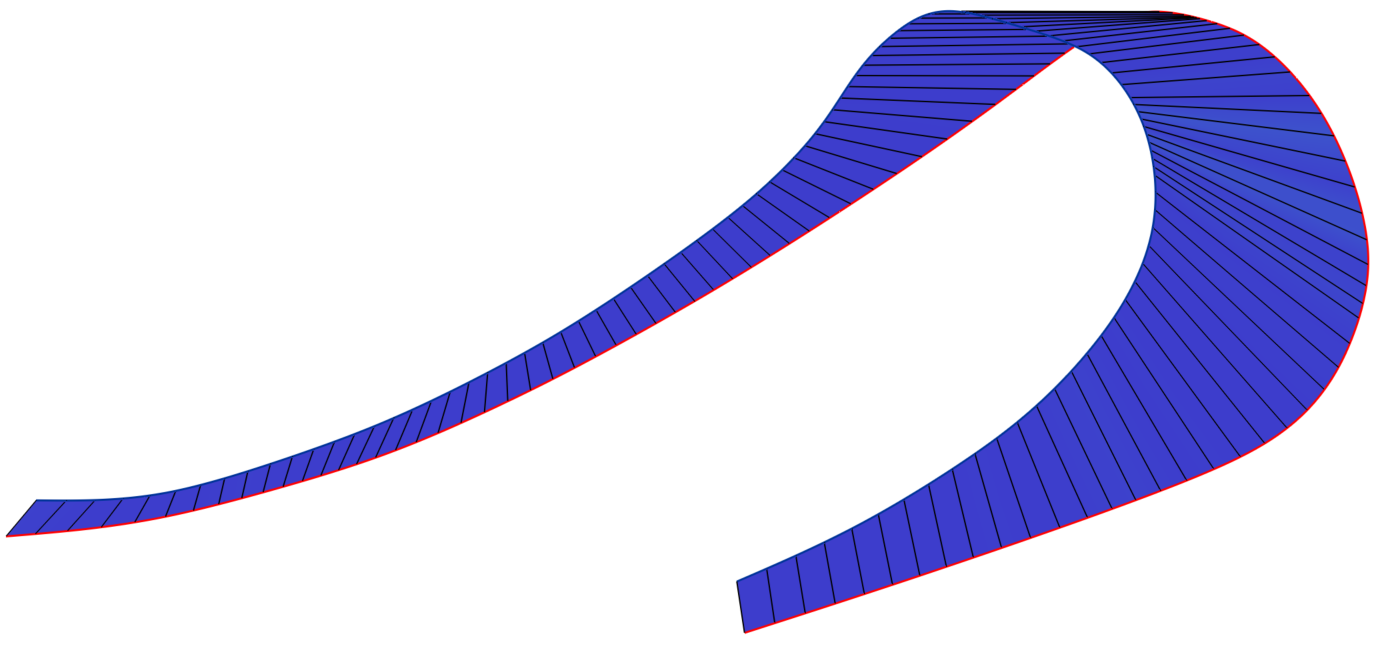}
}
\hfill
\subfigure[]{
\centering     
\includegraphics[width=0.25\textwidth]{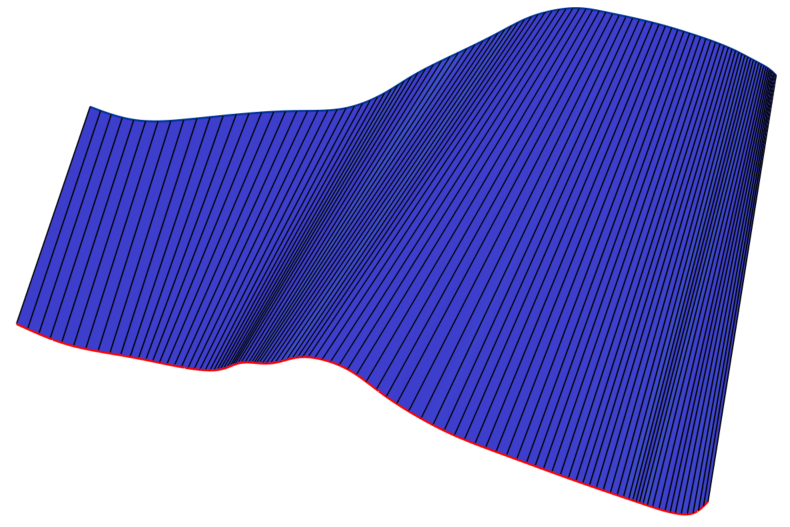}
}
}
\end{minipage}
\begin{minipage}[b]{0.9\linewidth}
\centerline{
\subfigure[]{
\centering     
\includegraphics[width=0.22\textwidth]{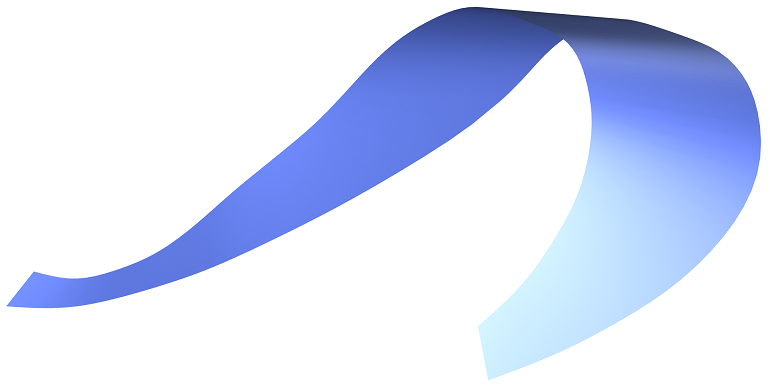}
}
\hfill
\subfigure[]{
\centering     
\includegraphics[width=0.25\textwidth]{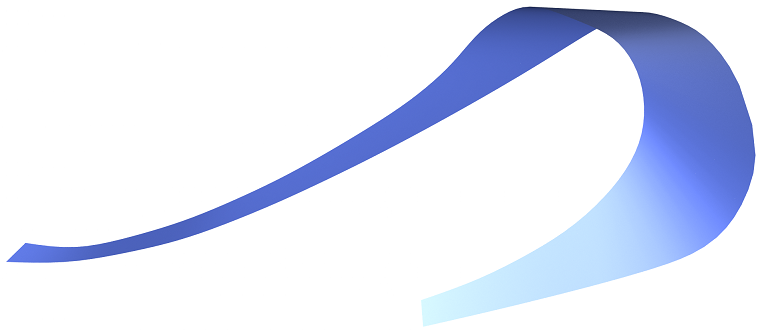}
}
\hfill
\subfigure[]{
\centering     
\includegraphics[width=0.25\textwidth]{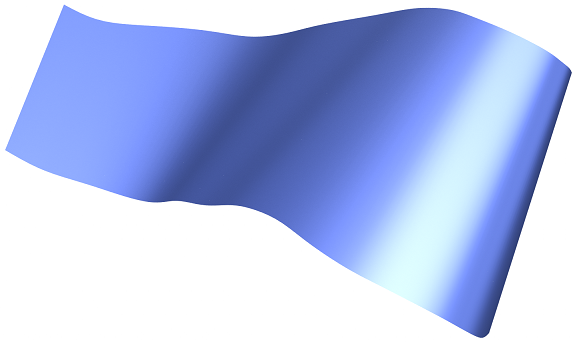}
}
}
\end{minipage}
\begin{minipage}[b]{0.9\linewidth}
\centerline{
\subfigure[]{
\centering     
\includegraphics[width=0.23\textwidth]{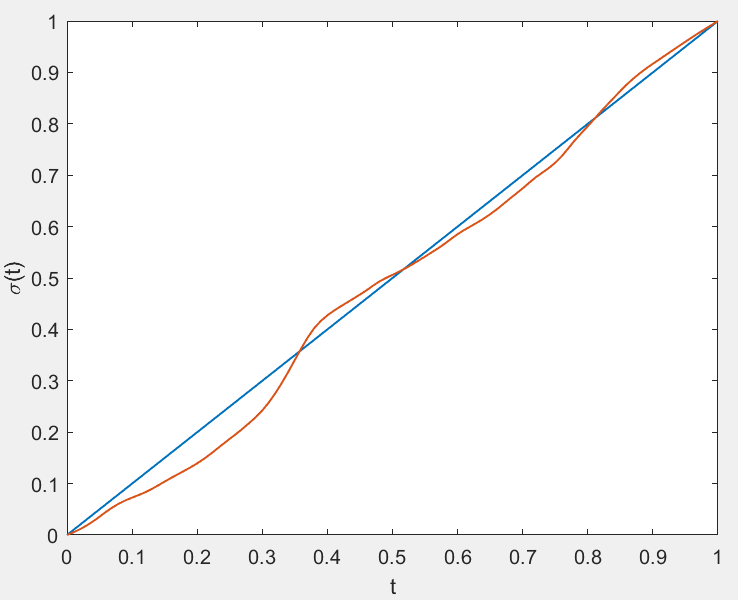}
}
\hfill
\subfigure[]{
\centering     
\includegraphics[width=0.23\textwidth]{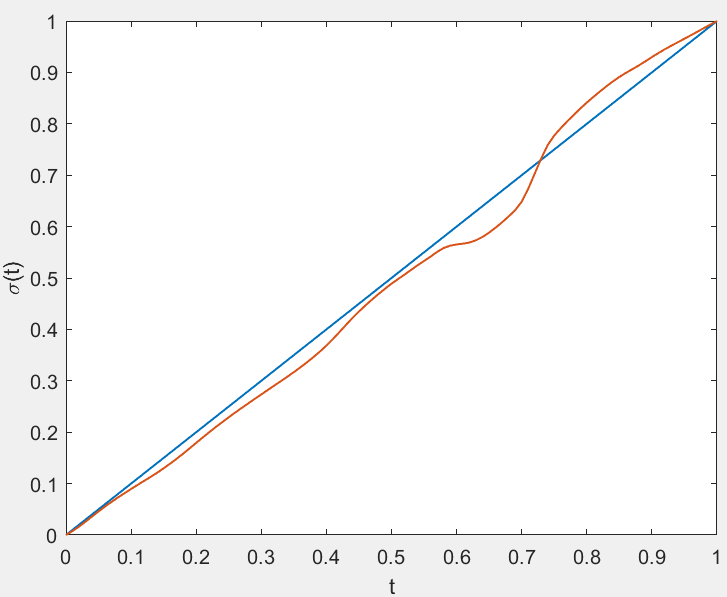}
}
\hfill
\subfigure[]{
\centering     
\includegraphics[width=0.23\textwidth]{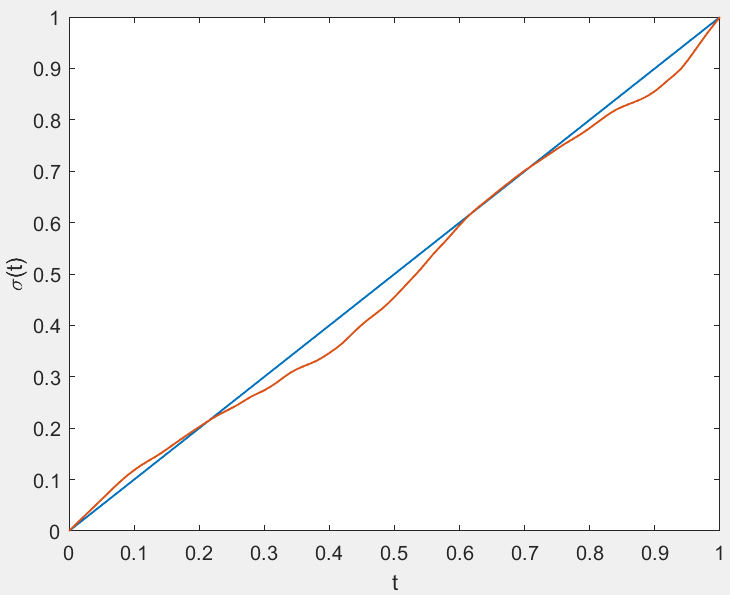}
}
}
\end{minipage}
\caption{A felt model. (a) A felt model with some specified curves on it. (b)-(d) Input curves for 3 patches. (e)-(g) Initial ruled surfaces whose (maximum,average) warp angles are (16.3\degree,4.2\degree), (30.1\degree,5.1\degree) and (17.8\degree,3.1\degree), respectively. (h)-(j) Resulting surfaces after optimization. The (maximum,average) warp angles are  (0.44\degree,0.07\degree), (1.17\degree,0.08\degree) and (0.50\degree,0.11\degree), respectively. (k)-(m) Rendered models. (n)-(p) Parametrization functions before and after optimization, where the blue curves refer to the optimized function and the red curves refer to the initial functions. For each strip, warp angles [0\degree, maximum degree of the initial ruled surface]  is mapped to  a color coding from blue to red.}
\label{fig:felt}
\end{figure}

\begin{figure}[tp]
\centering
\begin{minipage}[b]{0.92\linewidth}
\centerline{
\hfill
\subfigure[]{
\centering     
\includegraphics[width=0.3\textwidth]{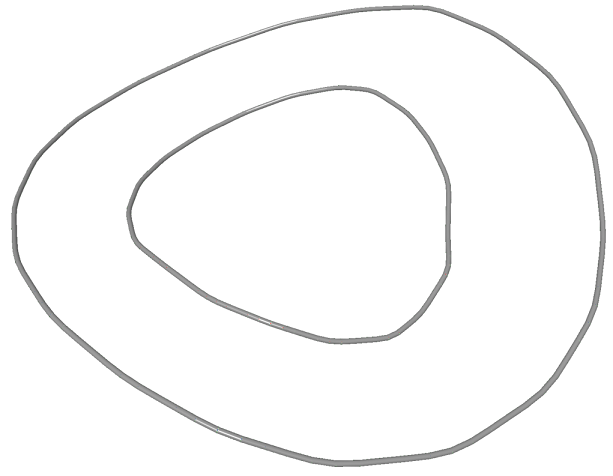}  
}
\subfigure[]{
\centering     
\includegraphics[width=0.3\textwidth]{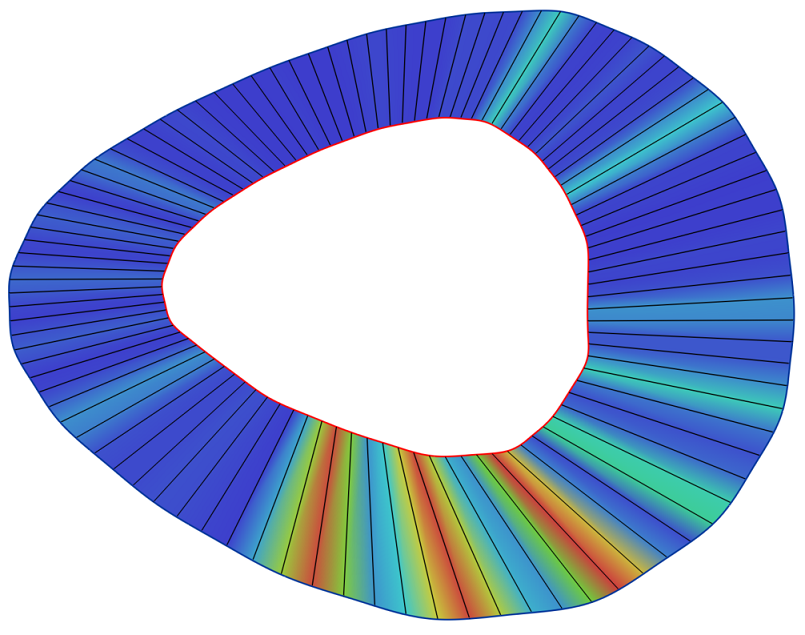}  
}
\hfill
\subfigure[]{
\centering     
\includegraphics[width=0.3\textwidth]{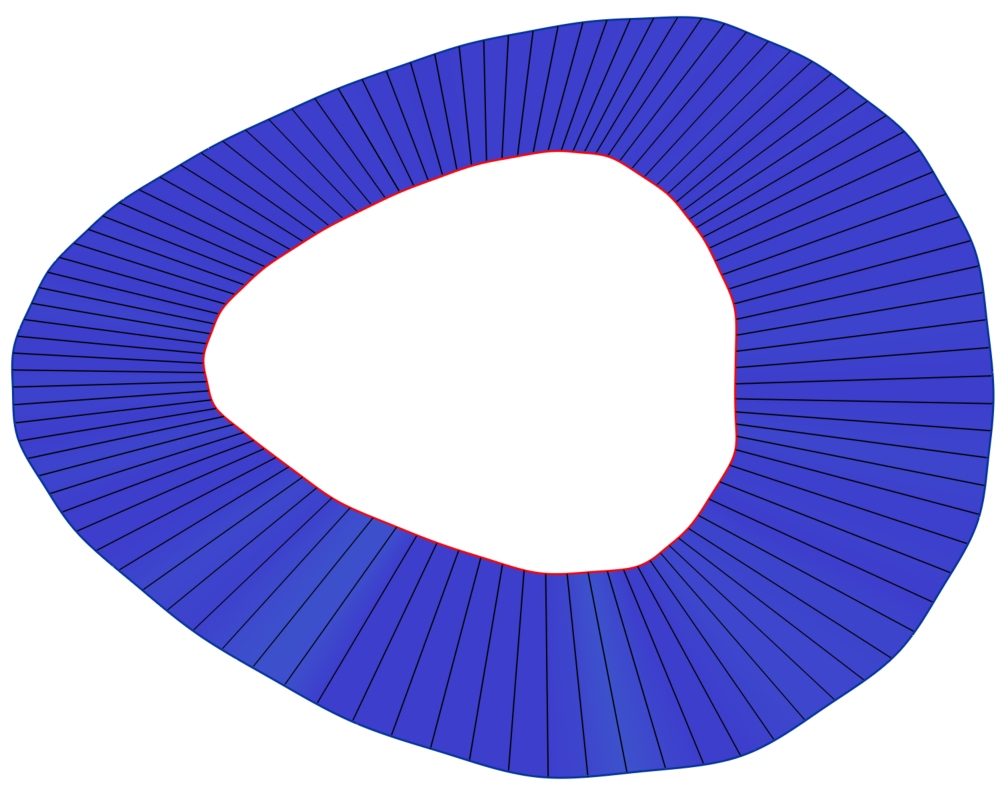}
}\hfill}
\end{minipage}
\vfill
\begin{minipage}[b]{0.92\linewidth}
\centerline{
\hfill
\subfigure[]{
\centering     
\includegraphics[width=0.3\textwidth]{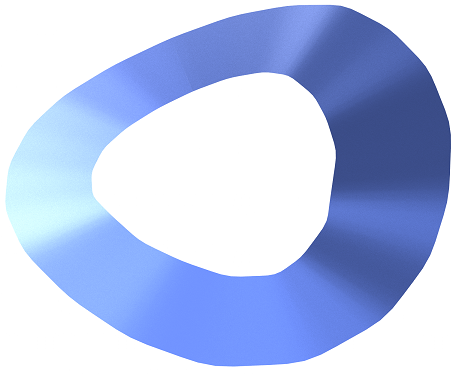}
}
\hfill
\subfigure[]{
\centering     
\includegraphics[width=0.3\textwidth]{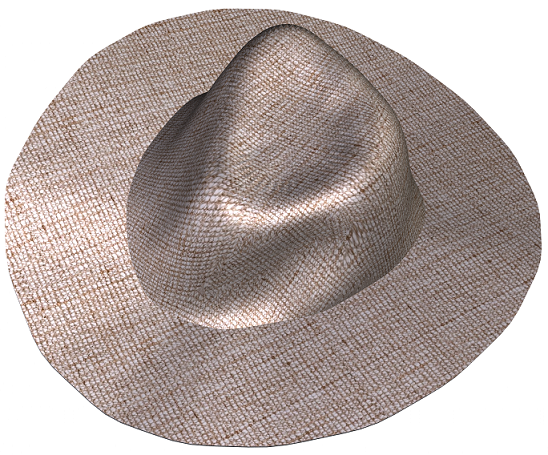}
}
\hfill
\subfigure[]{
\centering     
\includegraphics[width=0.3\textwidth]{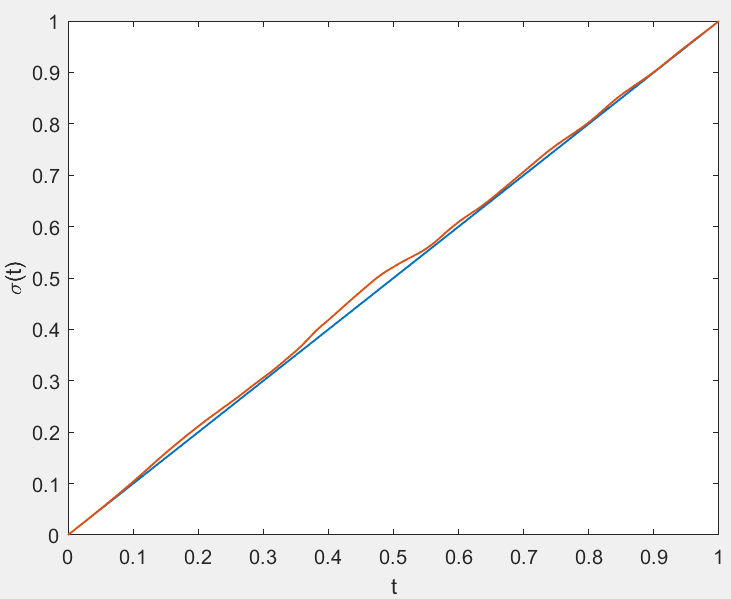}
}}
\end{minipage}
\caption{A hat brim model. (a) Input curves. (b) Initial ruled surface whose (maximum,average) warp angle is (11.4\degree,1.4\degree). (c) The resulting surface whose (maximum,average) warp angle is (0.44\degree,0.06\degree).  (d) Rendered strip.  (e) Rendered hat model with a texture. (f) Parametrization mapping functions before and after our optimization algorithm, where the blue curve refers to the optimized function and the red curve refers to the initial function.}
\label{fig:Hatstrip}
\end{figure}

\begin{figure}[!ht]
\centering
\begin{minipage}[b]{0.95\linewidth}
\centerline{
\hfill
\subfigure[]{
\centering     
\includegraphics[width=0.3\textwidth]{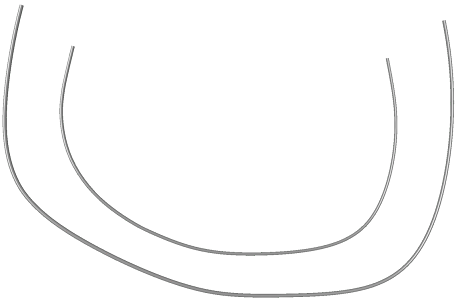}  
}
\subfigure[]{
\centering     
\includegraphics[width=0.3\textwidth]{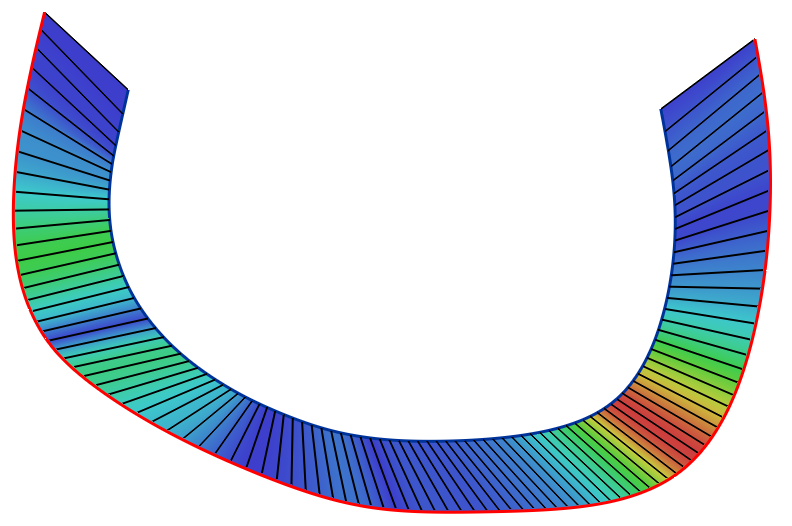}
}
\hfill
\subfigure[]{
\centering     
\includegraphics[width=0.3\textwidth]{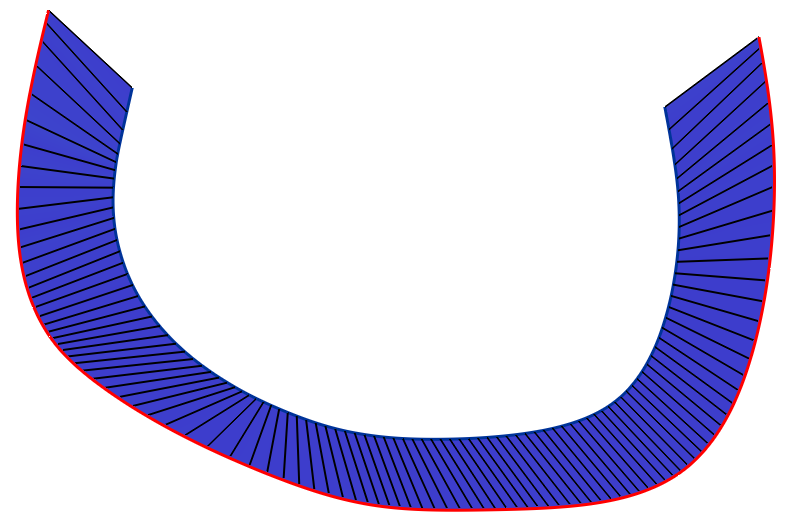}
}\hfill}
\end{minipage}
\vfill
\begin{minipage}[b]{0.95\linewidth}
\centerline{
\hfill
\subfigure[]{
\centering     
\includegraphics[width=0.25\textwidth]{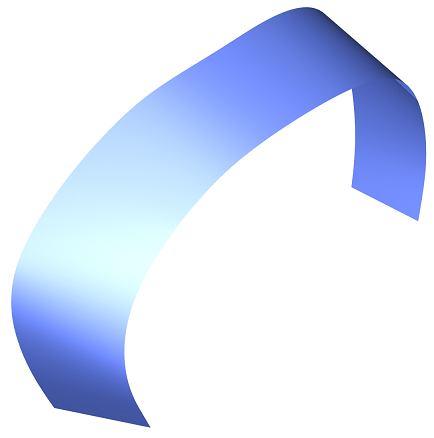}
}
\hfill
\subfigure[]{
\centering     
\includegraphics[width=0.25\textwidth]{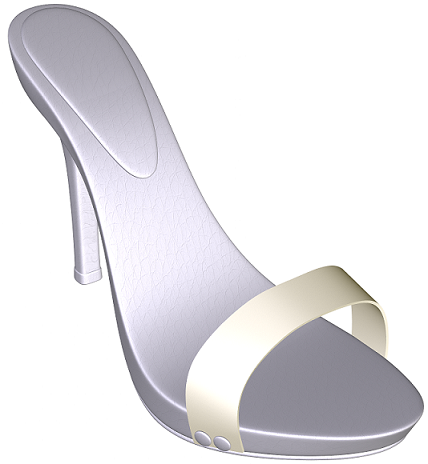}
}
\hfill
\subfigure[]{
\centering     
\includegraphics[width=0.3\textwidth]{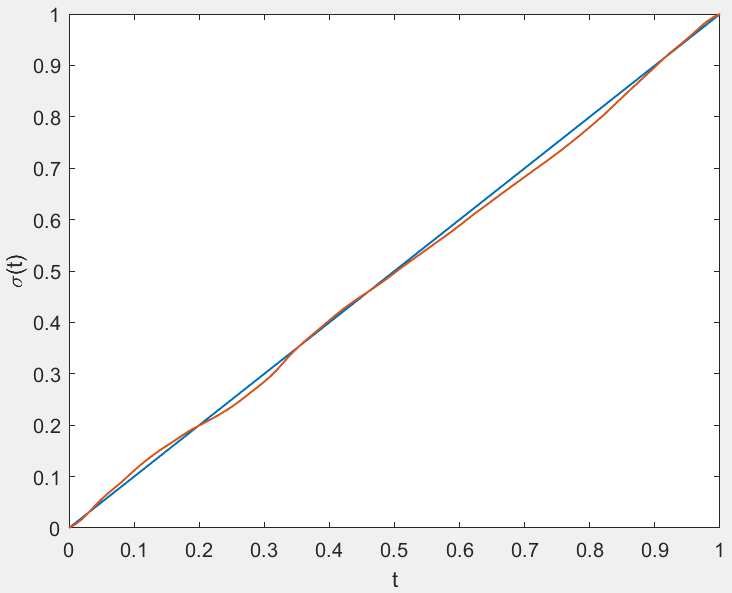}
}}
\end{minipage}
\caption{A shoe belt model. (a) Input curves. (b) Initial ruled surface whose (maximum,average) warp angle is (8.6\degree,2.1\degree). (c) The resulting surface obtained by out method whose  (maximum,average) warp angle is (0.11\degree,0.01\degree). (d) Rendered strip model. (e) Rendered model of a shoe with the resulting surface obtained with our method. (f) Parametrization functions before and after our optimization algorithm, where the blue curve refers to the optimized function and the red curve refers to the initial function.}
\label{fig:shoe}
\end{figure}

We show more examples in Figure ~\ref{fig:Ship-hull}, Figure ~\ref{fig:felt}, Figure ~\ref{fig:Hatstrip} and Figure ~\ref{fig:shoe}. In these experiments, we use a quadratic B-spline function for the parametrization mapping function with 50 control coefficients. We take 100 samples on a surface for evaluating surface developability in optimization. The computation time is around  1 second for all models.

 The  ship-hull models  in Figure ~\ref{fig:Ship-hull} are taken from ~\cite{Perez2007}. This boat model consists of 4 surface patches which are symmetric to a central plane. Therefore, computing 2 patches is enough for the design of the whole boat.    The input curves  (Figure ~\ref{fig:Ship-hull}(a), Figure ~\ref{fig:Ship-hull}(b)) are B-spline curves which are extended a bit in a pre-processing phase. The quasi-developable surfaces computed by the proposed algorithm  are trimmed by the lines connecting  the end points of  original input curves (Figure ~\ref{fig:Ship-hull}(g), Figure ~\ref{fig:Ship-hull}(h)) and are composed into a whole boat model shown in Figure ~\ref{fig:Ship-hull}(k), Figure ~\ref{fig:Ship-hull}(l). The graphics in Figure ~\ref{fig:Ship-hull}(i), Figure ~\ref{fig:Ship-hull}(j) shows the original and optimized parametrization mapping functions with the  inputs in Figure ~\ref{fig:Ship-hull}(a) and Figure ~\ref{fig:Ship-hull}(b), respectively.  The computation time are  0.4 seconds for both models in  Figure ~\ref{fig:Ship-hull}(a) and Figure ~\ref{fig:Ship-hull}(b).

The model in Figure ~\ref{fig:felt} is a felt model reconstructed from scanned data points, which is nearly developable.  We take 3  pairs of curves on the model and construct quasi-developable strips for each piece.  We show the original curves (Figure ~\ref{fig:felt} (b)-(d)), the initial ruled surface (Figure ~\ref{fig:felt} (e)-(g)), the optimized quasi-developable surface (Figure ~\ref{fig:felt} (h)-(j)) and  the rendered model (Figure ~\ref{fig:felt} (k)-(m)).  The original and optimized parametrization mapping functions are also shown for each surface strip (Figure ~\ref{fig:felt} (n)-(p)). The computation time for all three strips are about 0.3 seconds. 
 
The brim of the hat model in Figure ~\ref{fig:Hatstrip} is nearly a developable surface. We take two boundary curves and compute a quasi-developable surface using the proposed method. For this specific model, we fixed the end points of surface boundaries in optimization and obtain a  closed B-spline surface.  For end point fixing, an additional term measuring the distance from the end points of curves to its original positions is added into the objective function in Eqn.(\ref{eqn:objfun}) with a weight 1000. The computation time is about 0.5 seconds. 

The belt of the shoe model in Figure ~\ref{fig:shoe} is nearly developable. We take the boundary curves  of the belt as the input curves to our algorithm. The computation time is about 1.2 seconds.   We observe that the optimized mapping function are close to the original simple mapping function for the input curves in Figure ~\ref{fig:Hatstrip} and Figure ~\ref{fig:shoe}. Surface developability is, however, greatly improved  by optimization. This fact that surface developability is very sensitive to parametrization mapping is an evidence of the importance of a capability of continuous searching on input curves to explore the full solution space.

\subsection{Discussions and  limitations}
The experiments demonstrated the capability of using a  quadratic B-spline function with enough control coefficients for parametrization mapping. A linear B-spline function has a virtue that  it produces a resulting B-spline surface whose degree in the curved direction is the same as the degree of input B-spline curves.  Although using a linear B-spline mapping function, it is hard to achieve as good developability as using a quadratic B-spline function, surface developabilty obtained with a linear B-spline function can be fairly good for industrial applications. 

Because the lines connecting end points of input curves are in general not developable, the original input curves need to be extended  in some cases to obtain a surface with as large developability as possible. It is obvious that the shape of the extended curve has an impact on  surface developability. We extend the curve to an extra point  which is chosen relying on user experience, which can be improved by treating the extra points as   variables in optimization.

\section{Conclusions}

We have proposed an algorithm for computing a quasi-developable surface bounded by specified smooth curves as its boundary curves. Our method has a distinguish  advantage over existing methods in that it can explore the full solution space of smooth input curves and find a quasi-developable surface with large developability by solving a minimization problem. We also presented an algorithm to transform the resulting surface to a B-spline surface which is compatible with standard industrial representations of surfaces. 

For future works, we are going to study how to extend an input curve automatically so as to maximize surface developability of the extension region on a surface. We will also investigate how to perturb an input curve to expand the solution space of surfaces to achieve a larger developability.

\section*{Acknowledgements}
This work was partly supported by the National Natural Science Foundation of China (61672187) and Shandong Provincial Key Research and Development Project (2018GGX103038). The third author is partially supported by the National Natural Science Foundation of China (61602277) and Shandong Provincial Natural  Science Foundation of China (ZR2016FQ12).


\bibliography{developableMapping}

\end{document}